\documentclass[preprint,showpacs,amsfonts,aps,prc,floatfix,superscriptaddress]{revtex4} 
\usepackage{amsmath}
\usepackage{bm}
\usepackage{graphicx}

\begin{document}
\title{Low-mass dilepton production through transport process in quark-gluon plasma}
\date{\today}

\author{Yukinao Akamatsu}
\affiliation{Kobayashi-Maskawa Institute for the Origin of Particles and the Universe (KMI), Nagoya University, Nagoya 464-8602, Japan}
\affiliation{Department of Physics, The University of Tokyo, Tokyo 113-0033, Japan}

\author{Hideki Hamagaki}
\affiliation{Center of Nuclear Study, The University of Tokyo, Tokyo 113-0033, Japan}

\author{Tetsuo Hatsuda}
\affiliation{Theoretical Research Division, Nishina Center, RIKEN, Wako 351-0198, Japan}
\affiliation{Department of Physics, The University of Tokyo, Tokyo 113-0033, Japan}
\affiliation{Institute for the Physics and Mathematics of the Universe (IPMU), The University of Tokyo, Kashiwa 277-8568, Japan}

\author{Tetsufumi Hirano}
\affiliation{Department of Physics, The University of Tokyo, Tokyo 113-0033, Japan}
\affiliation{Department of Engineering and Applied Science, Sophia University, Tokyo 102-8554, Japan}

\begin{abstract}
We attempt to understand the low-mass dielectron enhancement observed by PHENIX Collaboration at Relativistic Heavy Ion Collider (RHIC) by transport peak in the spectral function.
On the basis of the second-order formalism of relativistic dissipative hydrodynamics, we parameterize the spectral function in low-frequency and long-wavelength region by two transport coefficients, electric diffusion coefficient $D$ and relaxation time $\tau_{\rm J}$, and compared our theoretical dielectron spectra with the experimental data.
We study spectrum of dielectrons produced in relativistic heavy ion collisions by using the profile of matter evolution under full (3+1)-dimensional hydrodynamics.
We find that the experimental data require the diffusion coefficient to be $D\geq 2/T$, with $T$ being temperature.
Our analysis shows that dielectrons emitted through transport process mainly come from high-temperature QGP phase.
\end{abstract}


\maketitle

\section{Introduction}
\label{DL:sec:1}

Study of the quark-gluon plasma (QGP) is currently undertaken by Relativistic Heavy Ion Collider (RHIC) at BNL and by Large Hadron Collider (LHC) at CERN.
At RHIC, signatures of the strongly correlated QGP have been accumulated through the analyses of the collective evolution of hot QCD matter \cite{Kolb:2003dz,Huovinen:2006jp,Hirano:2008hy} and of the energy loss of jets and heavy-quarks  \cite{Gyulassy:2003mc, Horowitz:2010yi}.
Electromagnetic probes, \textit{i.e.} photons and dileptons, are also important probes of the QCD matter since they are not contaminated by strong interaction and reflect transport property of the matter. 

Recently, PHENIX Collaboration at RHIC reported  enhancement of dielectrons in low-mass ($0.1 < m_{\rm ee} < 0.75$ GeV) region in Au+Au collisions \cite{:2007xw,Adare:2009qk}.
However, theoretical models, which were successful in reproducing the low-mass dilepton spectra at SPS (Super Proton Synchrotron), cannot explain the PHENIX data. 
This indicates the existence of yet unknown sources beyond the standard thermal radiations \cite{Rapp:2010sj,Ghosh:2010wt,Dusling:2007su,Dusling:2009ej,Bratkovskaya:2008bf,Bratkovskaya:2010gh}.
By modeling the spectral function based on the two scenarios discussed in the literature, the dropping mass and the width broadening, and combining it with the full (3+1)-dimensional hydrodynamic evolution, we also find that our model spectral function cannot explain the low-mass enhancement at PHENIX. 
(See Appendix \ref{DL:app:1} for details.)
As pointed out in \cite{Dusling:2007su}, one of the possible candidates which have tendency to fill the gap could be the processes considered in the Landau-Pomeranchuk-Migdal resummation, such as the off-shell annihilation process $q+q+\bar q\rightarrow q+\gamma^*\rightarrow q+{\rm e}^++{\rm e}^-$ \cite{Aurenche:2002wq}.
However the use of perturbative picture near transition temperature is not necessarily justified.

The main purpose of this paper is to study transport peak in the spectral function
and its consequence on dilepton spectra in low mass region.
Since the transport peak gives divergent dielectron rate in low-frequency limit $\omega\rightarrow 0$, this could be a possible source of the low-mass enhancement.
The transport peak reflects transport property of the QCD matter and has not been
fully considered in the above calculations.
 Perturbative calculations of the  transport coefficients and spectral function in low-frequency region \cite{Jeon:1994if,Arnold:2000dr,Arnold:2003zc,Hong:2010at,Hidaka:2010gh,Aarts:2002cc} require higher-order resummation and thus the transport peak originates from 
 various multiple scattering processes. Instead of taking the perturbative result,
we parametrize the transport peak at low-frequency and long-wavelength by a set of transport coefficients, the electric charge diffusion coefficient $D$ and relaxation time $\tau_{\rm J}$ \cite{TFT,Kadanoff}. Then we try to constrain their values by analyzing the dielectron data at PHENIX with the use of the full (3+1)-dimensional hydrodynamics simulation and the state-of-the-art lattice equation of state.
Furthermore, we compare the resultant constraint with the  perturbative QCD estimate at 
weak-coupling \cite{Arnold:2000dr,Arnold:2003zc,Hong:2010at} and the AdS/CFT estimate at strong coupling \cite{Natsuume:2007ty}.
We find that the main source of the low-mass dielectrons emitted through the transport process is the high-temperature QGP, not the low-temperature hadronic phase.

In Sec.~\ref{DL:sec:2}, we review the basics of the relativistic hydrodynamic model.
In Sec.~\ref{DL:sec:3}, we utilize the second-order formalism of relativistic dissipative hydrodynamics in the presence of external electromagnetic field to obtain the spectral function parametrized with two transport coefficients, $D$ and $\tau_{\rm J}$. 
Then, we calculate dielectron production using the spectral function with a transport peak and compare the  results with the experimental data.
In Sec.~\ref{DL:sec:4}, we give conclusion and outlook.

%%%%%%%%%%%%%%%%%%%%%%%%%%%%%%%%%%%%%%%%%%%%%%%%%%%%%%%%%%
\section{Relativistic hydrodynamics}
\label{DL:sec:2}

The relativistic hydrodynamic model has been quite successful in describing collective flow phenomena in heavy-ion collisions at RHIC \cite{Kolb:2003dz,Huovinen:2006jp,Hirano:2008hy}.
Its basic equation for perfect fluids reads 
\begin{eqnarray}
\label{DL:eq:hydro}
\partial_{\mu}T^{\mu\nu}=0, \ \ \ 
T^{\mu\nu}=(e+P)u^{\mu}u^{\nu}-Pg^{\mu\nu},
\end{eqnarray}
where $T^{\mu\nu}$ is the energy-momentum tensor, $e$ the energy density, $P$ the pressure, and $u^{\mu}$ the fluid velocity.
The baryon chemical potential is neglected, since it is small near mid-rapidity at RHIC and LHC energies.

\subsection{Lattice equation of state and spacetime evolution}
\label{DL:sec:2-a}

\begin{figure}
\begin{center}
\centering
\includegraphics[width=7cm, clip]{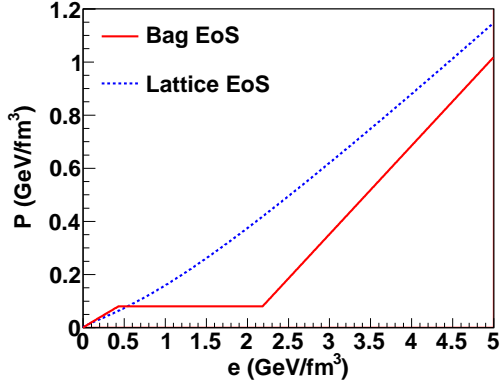}
\end{center}
\caption{
\footnotesize
(Color online)
Pressure as a function of energy density $P=P(e)$ compared with the conventional bag equation of state and the lattice equation of state \cite{Borsanyi:2010cj}.
In the bag model at $T=170$ MeV, pressure with $N_f = 3$ massless free ideal QGP gas is 
matched to that of a hadron resonance gas including resonances up to $\Delta(1232)$.
The resultant bag constant is $B^{1/4} = 247$ MeV.
}
\label{DL:fig:eos}
\end{figure}

The energy density and pressure are related through the equation of state (EoS) $P=P(e)$.
In the present paper, we use one of the latest  EoS obtained from (2+1)-flavor lattice QCD simulations with a Symanzik improved gauge action and a stout-link improved staggered fermion action \cite{Borsanyi:2010cj}.
The parameterization of the trace anomaly $I\equiv e-3P$ given in \cite{Borsanyi:2010cj} reads
\begin{eqnarray}
\label{DL:eq:LEOS} 
\frac{I(T)}{T^4}=\exp(-h_1/t-h_2/t^2)\cdot\left(h_0+\frac{f_0\cdot\left[\tanh(f_1\cdot t+f_2)+1\right]}{1+g_1\cdot t+g_2\cdot t^2}\right),
\end{eqnarray}
where $t\equiv T/(0.2 {\rm GeV})$, 
$(h_0, h_1, h_2)=(0.1396, -0.1800, 0.0350)$, 
$(f_0, f_1, f_2)=(2.76, 6.79, -5.29)$, and 
$(g_1, g_2)=(-0.47, 1.04)$.
As shown in Fig.~\ref{DL:fig:eos}, the lattice EoS shows a smooth crossover from the hadronic phase to the QGP phase in contrast to the historic bag EoS, which has the first order phase transition. 

In solving hydrodynamic equation Eq.~(\ref{DL:eq:hydro}), we pose the initial condition at $\tau_0=0.6$ fm/$c$ for entropy density $s$ and flow vector $\vec u$:
\begin{eqnarray}
\label{ST:eq:initial}
\tau_0 s(\eta_{\rm s},\vec x_{\bot})&=&C\cdot\theta(y_{\rm b}-|\eta_{\rm s}|)\nonumber \\
&&\times f^{\rm pp}(\eta_{\rm s})
\left[a \left(
\frac{y_{\rm b}-\eta_{\rm s}}{y_{\rm b}}\frac{dN^{\rm A}_{\rm part}}{d^2x_{\bot}}
+\frac{y_{\rm b}+\eta_{\rm s}}{y_{\rm b}}\frac{dN^{\rm B}_{\rm part}}{d^2x_{\bot}}\right)
+(1-a)\frac{dN_{\rm coll}}{d^2x_{\bot}}\right], \\ 
u_z&=&\sinh \eta_{\rm s}, \ \ u_x=u_y=0,
\end{eqnarray}
when nucleus A(B) is traveling along $z$-axis to negative (positive) direction.
Here $y_{\rm b} \ (>0)$ is the beam rapidity, $C=13.0$ and $a=0.85$ are fitting parameters, $dN^{\rm A(B)}_{\rm part}/d^2x_{\bot},dN_{\rm coll}/d^2x_{\bot}$ are defined in the Glauber model \cite{Miller:2007ri}, and $f^{\rm pp}(\eta_{\rm s})$ is a parametrization of the shape of rapidity distribution in proton-proton (p+p) collision at $\sqrt{s_{\rm NN}}=200$ GeV.
For details of the initial condition, see Refs.~\cite{Hirano:2001eu,Hirano:2002ds,Hirano:2005xf,Hirano:2007ei}.
With this initial condition, hydrodynamic equations Eq.~(\ref{DL:eq:hydro}) are numerically solved assuming chemical equilibrium above freezeout temperature $T_{\rm f}$, which we determine as follows.

\begin{figure}
\centering
\includegraphics[width=5cm, angle=-90,clip]{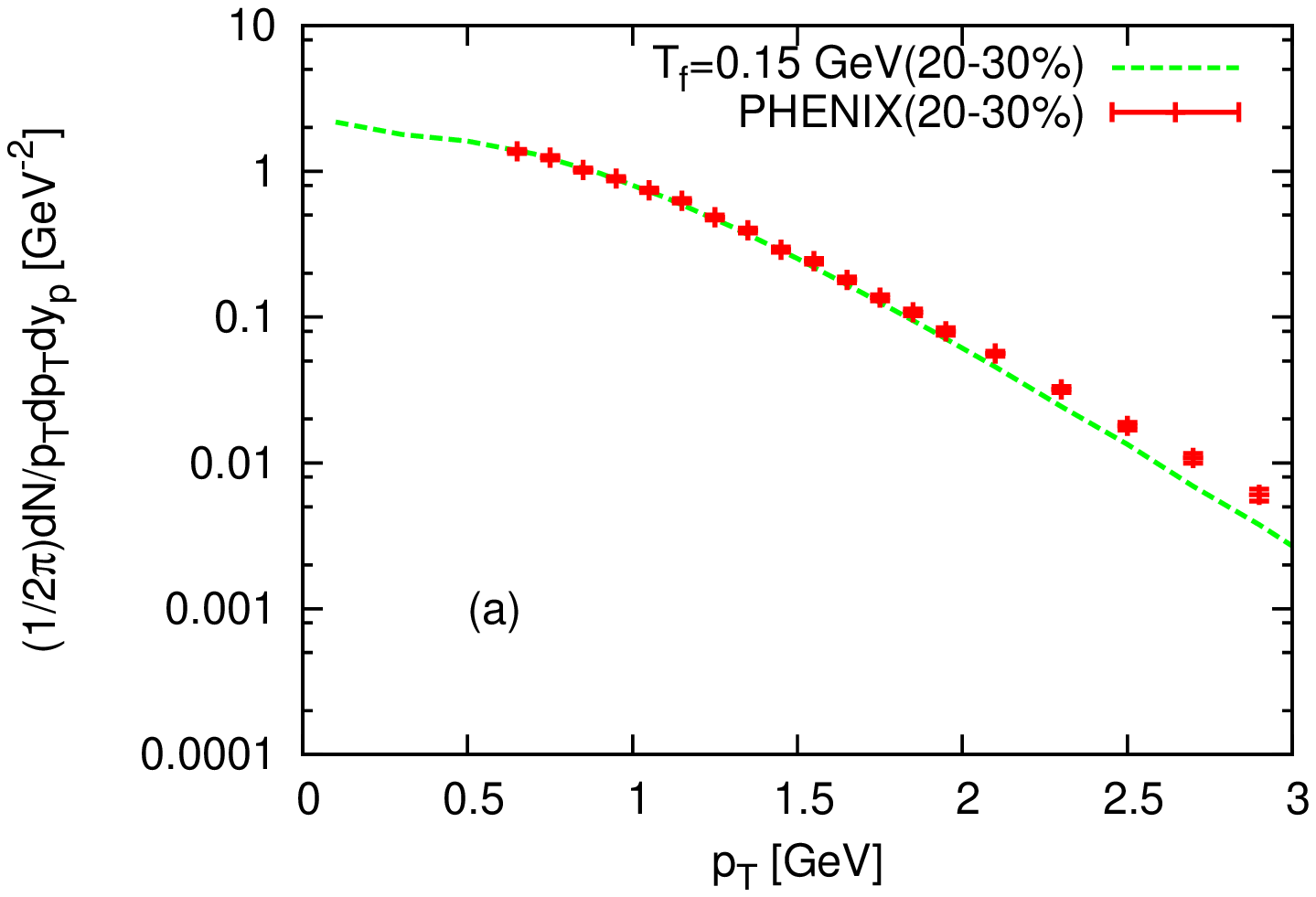}
\includegraphics[width=5cm, angle=-90,clip]{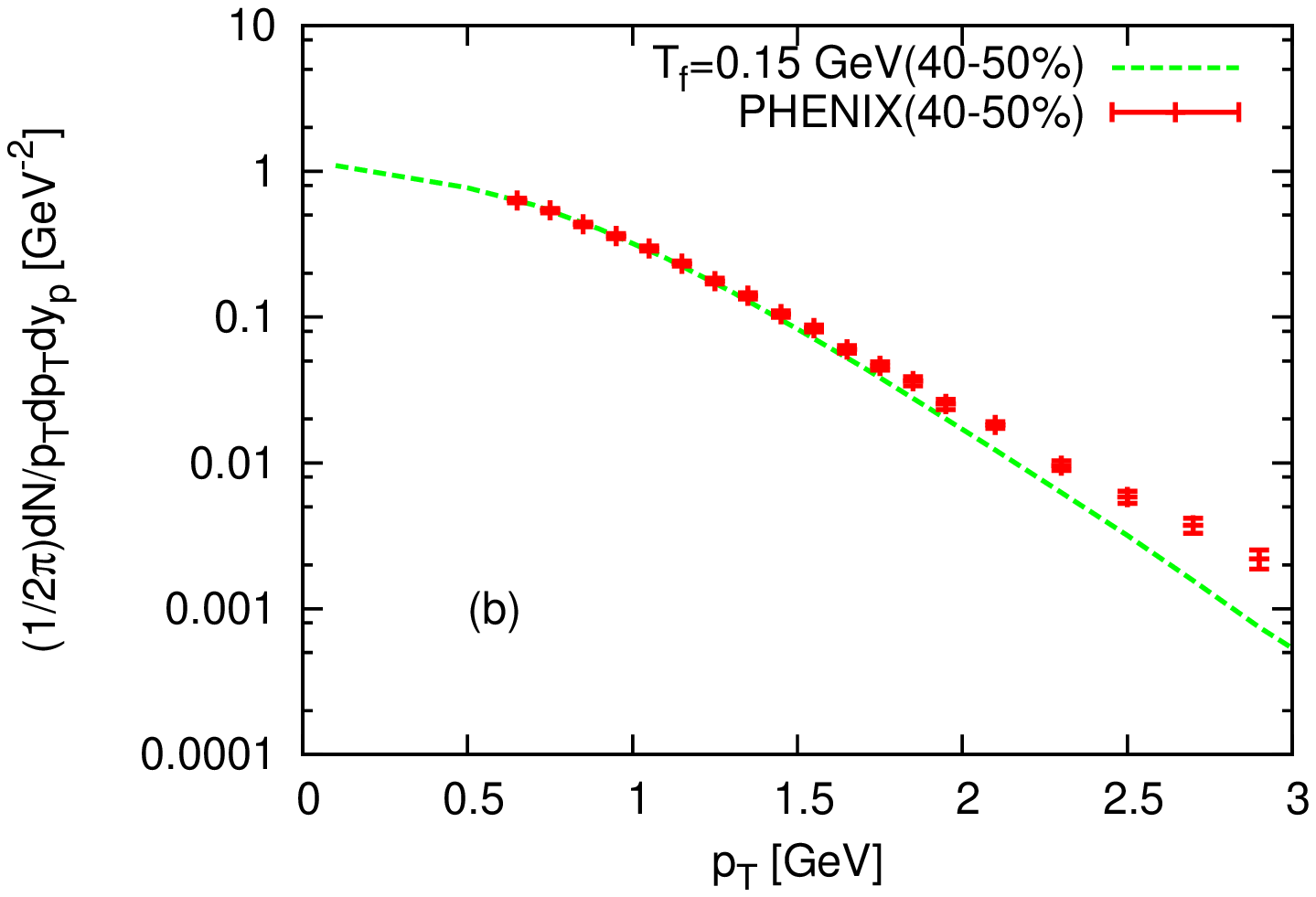}
\caption{
\footnotesize
(Color online)
Shown is the hydrodynamic calculation for the slope of $p_{_{\rm T}}$ spectra of protons with freezeout temperature $T_{\rm f}=0.15$ GeV.
The $p_{_{\rm T}}$ spectra are calculated by the hydrodynamic model with lattice EoS.
In the hydrodynamic calculation, chemical equilibrium is assumed at $T>T_{\rm f}=0.15$ GeV.
The impact parameters of the collision are (a) $b=7.1$ fm (20-30\% centrality) and (b) $b=9.7$ fm (40-50\% centrality) in the hydrodynamic calculation.
Note that overall $p_{_{\rm T}}$ spectra are scaled (by multiplying $\approx 1.4$) to match the experimental data at $p_{_{\rm T}}=1.0$ GeV.
}
\label{DL:fig:Tf}
\end{figure}

At $T=T_{\rm f}$, Cooper-Frye formula \cite{Cooper:1974mv} converts the profile of hydrodynamic evolution $(\vec u, T)$ into spectra of particles and resonances (up to $\Delta(1232)$) by
\begin{eqnarray}
\label{ST:eq:Cooper-Frye}
E\frac{dN_{i}}{d^{3}p}=\frac{d_{i}}{(2\pi)^{3}}\int_{\partial\Sigma}
\frac{p^{\mu}d\sigma_{\mu}}{\exp\left[p^{\mu}u_{\mu}/T_{\rm f}\right]\mp 1},
\end{eqnarray}
where $d_{i}$ is the degeneracy factor of hadron species $i$, the sign in the denominator is $-$ ($+$) for bosons (fermions), $\partial\Sigma$ is the spacetime hypersurface satisfying $T(\tau,\eta_{\rm s},x,y)=T_{\rm f}$, and $d\sigma^{\mu}$ is its element.
To obtain spectra of stable particles, the decay products from $\Delta(1232)$ are also included.
We determine the thermal freezeout temperature $T_{\rm f}$ by fitting the experimental slope of the proton $p_{_{\rm T}}$ spectra as shown in Fig.~\ref{DL:fig:Tf}. 
The slopes can be fitted well by $T_{\rm f}=0.15$ GeV.
The impact parameters of the simulation are $b=7.1$ fm (20-30\% centrality) and $b=9.7$ fm (40-50\% centrality).
Note that we have multiplied a factor $\approx 1.4$ to fit the absolute magnitude of the experimental data at $p_{_{\rm T}}=1.0$ GeV.
This slight difference in the magnitude between hydrodynamical calculation and experimental data is possibly due to an absence of chemical freezeout mechanism \cite{Hirano:2002ds} in the lattice equation of state.

\subsection{Dielectron spectrum}
\label{DL:sec:2-b}

Here we briefly summarize the formula for the dielectron production from 
expanding medium in relativistic heavy-ion collisions.
Since the invariant mass of our interest is $\sqrt{q^2}\geq 0.1$ GeV, we neglect the electron mass in the following.
The (3+1)-dimensional hydrodynamic model gives each spacetime point a local flow vector $u^{\mu}(x)$ and a local temperature $T(x)$.
Each spacetime volume under hydrodynamic evolution emits lepton pairs with the rate \cite{Feinberg:1976ua,McLerran:1984ay}
\begin{eqnarray}
\label{DL:eq:formula}
\frac{E_1E_2dN_{l^+l^-}}{d^3p_1d^3p_2d^4x}
&=&\frac{\alpha^2}{2\pi^4q^4}\frac{p_1^{\mu}p_2^{\nu}+p_2^{\mu}p_1^{\nu}-\frac{q^2}{2}g^{\mu\nu}}{\exp (q^0/T)-1}
{\rm Im}G_{\rm R}^{\mu\nu}(q;T),\\
G_{\rm R}^{\mu\nu}(q;T)&\equiv& \int d^4x e^{iqx}i\theta(x^0)\langle[J^{\mu}(x),J^{\nu}(0)]\rangle_T,\\
J^{\mu}(x)&\equiv&
\frac{2}{3}\bar {\rm u}\gamma^{\mu}{\rm u}
-\frac{1}{3}\bar {\rm d}\gamma^{\mu}{\rm d}
-\frac{1}{3}\bar {\rm s}\gamma^{\mu}{\rm s},
\end{eqnarray}
when observed in the local fluid rest frame.
In the laboratory frame, momentum distribution of the lepton pairs emitted from the thermal medium in heavy ion collisions is expressed by the following integration:
\begin{eqnarray}
\label{DL:eq:formula_hydro}
\frac{E_1E_2dN}{d^3p_1d^3p_2}
&=&\int_{\Sigma} d^4x\frac{\alpha^2}{2\pi^4\tilde q^4}\frac{\tilde p_1^{\mu}\tilde p_2^{\nu}+\tilde p_2^{\mu}\tilde p_1^{\nu}-\frac{\tilde q^2}{2}g^{\mu\nu}}{\exp (\tilde q^0/T(x))-1}
{\rm Im}\tilde G_{\rm R}^{\mu\nu}(\tilde q;T(x)).
\end{eqnarray}
$\Sigma$ indicates the spacetime region of the hot QCD medium with $T>T_{\rm f}$ which is considered to radiate dielectrons.
The $\tilde A^{\mu\nu\cdots}$ denotes a Lorentz tensor in the local rest frame boosted by $u^\mu$ from that in the laboratory frame $A^{\mu\nu\cdots}$.

In the PHENIX experiment, with which we will compare our results later, the acceptance of each electron or positron depends on the detector geometry and on the particle kinematics \cite{:2007xw,Adare:2009qk}.
We introduce an acceptance function $A(p_1,p_2)$ to pick up the phase space where both the electron and positron are in the acceptance.
The dielectron invariant mass spectrum from the expanding medium is therefore given by
\begin{eqnarray}
\label{DL:eq:formula_acc}
\frac{dN}{dm_{\rm ee}}&=&2m_{\rm ee}\int d^3p_1d^3p_2\frac{dN}{d^3p_1d^3p_2}
A(p_1,p_2)\delta(q^2-m_{\rm ee}^2).
\end{eqnarray}

%%%%%%%%%%%%%%%%%%%%%%%%%%%%%%%%%%%%%%%%%%%%%%%%%%%%%%%%%%
\section{Model of Low-mass dileptons: transport peak}
\label{DL:sec:3}

One of the promising non-perturbative approaches to explain the low-mass enhancement is to utilize the transport theory, which parameterizes the spectral function in low-energy and long-wavelength region with transport coefficients.

\subsection{Linearized hydrodynamics with external field}
\label{DL:sec:3-a}

Here we construct retarded correlator $G_{\rm R}^{\mu \nu}(q;T)$ and corresponding spectral function (transport-SPF)  from the linear analysis of transport equations.
According to the linear response theory, the retarded correlator relates small perturbation $\delta H(t)=\int d^3x J^{\mu}(x)\delta A_{\mu}(x)$ and system response $\langle \delta J^{\mu}(q)\rangle_T$ in the linear order \cite{TFT}:
\begin{eqnarray}
\label{DL:eq:linear_response}
\langle \delta J^{\mu}(q)\rangle_T=-G_{\rm R}^{\mu\nu}(q;T)\delta A_{\nu}(q),
\end{eqnarray}
where $f(q)\equiv \int d^4x e^{iqx}f(x)$.
Hereafter we calculate the linear relation between $\delta J^{\mu}$ and $\delta A_{\nu}$ on the basis of relativistic dissipative hydrodynamics.

In the presence of external field $\delta A^{\mu}$, conservation laws for the energy-momentum tensor $T^{\mu\nu}$ and the electric current $J^{\mu}$ are modified to
\begin{eqnarray}
\label{DL:eq:conservation1}
\partial_{\nu}T^{\nu\mu}&=&F^{\mu\nu}J_{\nu},\\
\label{DL:eq:conservation2}
\partial_{\mu}J^{\mu}&=&0,
\end{eqnarray}
where we define $F_{\mu\nu}\equiv \partial_{\mu}\delta A_{\nu}-\partial_{\nu}\delta A_{\mu}$, $\vec E\equiv -\vec\nabla \delta A^0-\frac{\partial\delta\vec A}{\partial t}$, and $\vec B\equiv\vec\nabla\times\delta\vec A$.
In relativistic viscous hydrodynamics in Landau frame \cite{Landau, footnote1}, the energy-momentum tensor $T^{\mu\nu}$ and the electric current $J^{\mu}$ are decomposed as
\begin{eqnarray}
T^{\mu\nu}&=&eu^{\mu}u^{\nu}-(P+\Pi)\triangle^{\mu\nu}+\pi^{\mu\nu},\\
\label{DL:eq:viscous_hydro}
J^{\mu}&=&\rho u^{\mu}+\nu^{\mu},\\
\triangle^{\mu\nu}&\equiv& g^{\mu\nu}-u^{\mu}u^{\nu},
\end{eqnarray}
with bulk pressure $\Pi$, shear stress tensor $\pi^{\mu\nu}$, and dissipative electric current $\nu^{\mu}$ satisfying $\pi^{\mu\nu}u_{\nu}=0,\pi^{\mu}_{\mu}=0,\nu^{\mu}u_{\mu}=0$.
The scalar quantities $e,P$, and $\rho$ satisfy the equation of state $P=P(e,\rho )$ derived in equilibrium.
The entropy current in the second-order formalism constructed by Israel and Stewart \cite{Israel:1976tn,Israel:1979wp} is decomposed as
\begin{eqnarray}
s^{\mu}=su^{\mu}-\frac{\mu}{T}\nu^{\mu}
-\frac{1}{T}(\alpha_0\Pi\nu^{\mu}+\alpha_1\pi^{\mu\nu}\nu_{\nu})
-\frac{u^{\mu}}{2T}(\beta_0\Pi^2 -\beta_1\nu^{\mu}\nu_{\mu}+\beta_2\pi^{\rho\sigma}\pi_{\rho\sigma}),
\end{eqnarray}
with positive transport coefficients $\alpha_a$ $(a=0,1)$ and $\beta_b$ $(b=0,1,2)$ and chemical potential $\mu$ for the electric charge.
Here $s$ is the entropy density $s(e,\rho)$ derived from the equation of state.
Calculating the divergence of the entropy current up to quadratic order in $\Pi,\pi^{\mu\nu},\nu^{\mu}$, and $F^{\mu\nu}$ and in derivatives of $e,\rho$, and $u^{\mu}$, we obtain
\begin{eqnarray}
\label{DL:eq:entropy_production}
\partial_{\mu}s^{\mu}&=&
-\frac{\Pi}{T}(\partial_{\mu}u^{\mu}+\alpha_0\partial_{\mu}\nu^{\mu}+\beta_0\dot\Pi)
+\frac{\pi^{\mu\nu}}{T}(\partial_{\mu}u_{\nu}-\alpha_1\partial_{\mu}\nu_{\nu}-\beta_2\dot\pi_{\mu\nu})\nonumber\\
&&-\frac{\nu^{\mu}}{T}\left[T\partial_{\mu}\left(\frac{\mu}{T}\right)+F_{\mu\nu}u^{\nu}+\alpha_0\partial_{\mu}\Pi+\alpha_1\partial_{\nu}\pi^{\nu}_{\mu}-\beta_1\dot\nu_{\mu}\right],
\end{eqnarray}
where $\dot f\equiv u^{\mu}\partial_{\mu}f$.
The second law of thermodynamics requires the following constitutive equations:
\begin{eqnarray}
-\Pi&=&\zeta(\partial_{\mu}u^{\mu}+\alpha_0\partial_{\mu}\nu^{\mu}+\beta_0\dot\Pi),\\
\pi_{\mu\nu}&=&2\eta\langle\langle\partial_{\mu}u_{\nu}
-\alpha_1\partial_{\mu}\nu_{\nu}
-\beta_2\dot\pi_{\mu\nu}\rangle\rangle,\\
\label{DL:eq:constitutive}
\nu^{\mu}&=&\sigma \triangle^{\mu\rho}\left[T\partial_{\rho}\left(\frac{\mu}{T}\right)+F_{\rho\sigma}u^{\sigma}
+\alpha_0\partial_{\rho}\Pi+\alpha_1\partial_{\sigma}\pi^{\sigma}_{\rho}
-\beta_1\dot\nu_{\rho}\right],
\end{eqnarray}
with bulk and shear viscosities $\zeta, \ \eta \ (\geq0)$ and electrical conductivity $\sigma \ (\geq 0)$.
Here $\langle\langle B^{\mu\nu}\rangle\rangle$ stands for a spatial, symmetric, and traceless tensor extracted from a general tensor $B^{\mu\nu}$:
\begin{eqnarray}
\langle\langle B^{\mu\nu}\rangle\rangle\
\equiv \triangle^{\mu\rho}\triangle^{\nu\sigma}
\left[\frac{B_{\rho\sigma}+B_{\sigma\rho}}{2}-\frac{\triangle_{\rho\sigma}\triangle^{\alpha\beta}B_{\alpha\beta}}{3}\right].
\end{eqnarray}

When the system is close to equilibrium, the energy density $e(x)$, the charge density $\rho (x)$ and the flow vector $u^{\mu}(x)$ slightly deviate from their equilibrium quantities:
\begin{eqnarray}
e(x)&=&e+\delta e(x),\\
\label{DL:eq:charge_density}
\rho (x)&=&\rho +\delta \rho (x),\\
\label{DL:eq:flow_vector}
u^{\mu}(x)&=&(1,\ \delta \vec u(x)).
\end{eqnarray}
Neglecting for simplicity the couplings $\alpha_{0,1}$ between different dissipative modes, the dissipative part of the electric current $\nu^{\mu}(x)$ is then given by
\begin{eqnarray}
\nu^0(x)&=&0, \\
\vec\nu(x)&=&-\sigma\left[
T\vec\nabla\left(\frac{\mu}{T}\right)
-\vec E+\beta_1\partial_t\vec\nu
\right],
\end{eqnarray}
in the linear order in $\delta e$, $\delta \rho$, $\delta \vec u$, and $A^{\mu}$.
Explicitly in terms of $\delta e$, and $\delta \rho$,
\begin{eqnarray}
T\vec\nabla\left(\frac{\mu}{T}\right)&=&
\frac{1}{X}
\left(\frac{\partial e}{\partial T}+\frac{\mu}{T}\frac{\partial e}{\partial \mu}\right)\vec\nabla\delta \rho
-\frac{1}{X}\left(\frac{\partial \rho}{\partial T}+\frac{\mu}{T}\frac{\partial \rho}{\partial\mu}\right)
\vec\nabla\delta e, \\
{X}&\equiv&
{\frac{\partial e}{\partial T}\frac{\partial \rho}{\partial \mu}
-\frac{\partial e}{\partial \mu}\frac{\partial \rho}{\partial T}}.
\end{eqnarray}
In the quark-gluon plasma with vanishing baryon density, $\rho=\frac{\partial e}{\partial \mu}=\frac{\partial \rho}{\partial T}=0$ holds and the electric current is obtained as
\begin{eqnarray}
J^{\mu}(x)&=&(\delta \rho (x),\vec\nu(x)),\\
\label{DL:eq:constitutive_linear}
\vec\nu(x)&=&\sigma\vec E
-D\vec\nabla\delta \rho
-\tau_{\rm J}\frac{\partial \vec\nu}{\partial t},\\
D&\equiv& \frac{\sigma}{\chi},\ 
\tau_{\rm J} \ \equiv \ \beta_1\sigma, \ 
\chi \ \equiv \ \frac{\partial \rho}{\partial \mu}.
\end{eqnarray}
The dynamics of the electric current $J^{\mu}(x)$ is obtained in a closed form, while with finite electric charge density there arises a coupling between the sound mode $\delta e(x)$ and the dissipative electric current $\nu^{\mu}(x)$ even after ignoring $\alpha_{0,1}$.
This is because energy current, or momentum, carries electric charge at finite density $\mu\neq 0$.

Combining Eqs.~(\ref{DL:eq:conservation2}), (\ref{DL:eq:viscous_hydro}), and (\ref{DL:eq:constitutive_linear}), we obtain
\begin{eqnarray}
J^0(q)&=&
\frac{\chi D\left(-k^2\delta A_0(q)-\omega k^i{\delta A_i(q)}\right)}{-\tau_{\rm J} \omega^2-i\omega+D k^2},\\
J^i(q)&=&
\frac{-\chi D \omega k^i}{-\tau_{\rm J} \omega^2-i\omega+Dk^2}
\left(
\delta A_0(q)
+\frac{\omega k^j\delta A_j(q)}{k^2}
\right)
\nonumber\\
&&+\frac{\chi D \omega}{-\tau_{\rm J}\omega-i}\left(
-\delta A_i(q)+\frac{k^ik^j\delta A_j(q)}{k^2}
\right),
\end{eqnarray}
where $q^{\mu}=(\omega,\vec k)$ and upper and lower Lorentz indices matter, $k_x=k^{i=1}=-k_{i=1}$ for example.
The retarded correlators are given by
\begin{eqnarray}
G_{\rm R}^{00}(\omega,\vec k;T)
&=&\frac{\sigma k^2}{-\tau_{\rm J} \omega^2-i\omega+D k^2},\\
G_{\rm R}^{0i}(\omega,\vec k;T)
&=& G_{\rm R}^{i0}(\omega,\vec k;T)
\ = \ \frac{\sigma \omega k^i}{-\tau_{\rm J} \omega^2-i\omega+D k^2},\\
G_{\rm R}^{ij}(\omega,\vec k;T)
&=&\frac{\sigma \omega}{-\tau_{\rm J}\omega-i}\left(\delta^{ij}-\frac{k^ik^j}{k^2}\right)
+\frac{\sigma\omega^2}{-\tau_{\rm J} \omega^2-i\omega+D k^2}\frac{k^ik^j}{k^2},
\end{eqnarray}
and we obtain two independent components of the spectral function
\begin{eqnarray}
\label{DL:eq:SPF_L}
{\rm Im}G^{\rm (L)}_{\rm R}(q;T)
&\equiv&-\frac{q^2}{k^2}{\rm Im}G^{00}_{\rm R}(q;T)
=-\frac{\chi D \omega q^2}{\omega^2+(\tau_{\rm J}\omega^2-Dk^2)^2},\\
\label{DL:eq:SPF_T}
{\rm Im}G^{\rm (T)}_{\rm R}(q;T)
&\equiv&\frac{1}{2}\left({\rm Im}G^{\mu}_{\rm R,\mu}(q;T)
-{\rm Im}G^{\rm (L)}_{\rm R}(q;T)\right)
=-\frac{\chi D\omega}{\tau_{\rm J}^2\omega^2+1}.
\end{eqnarray}

Here we have demonstrated our calculation in one-flavor case, but generalization to multi-flavor case can be achieved straightforwardly.
We discuss in Appendix~\ref{DL:app:2} that the transport equation Eq.~(\ref{DL:eq:constitutive_linear}) holds in 3-flavor case under three assumptions: 
(i) There is no coupling between the dissipative modes in the decomposition of the entropy current,
(ii) Transport coefficients are flavor-independent, and
(iii) Quark number susceptibility matrix is proportional to the unit matrix $\chi_{ij}\propto \delta_{ij}$.
Here $\chi_{ij}\equiv\frac{\partial n_i}{\partial \mu_j}$ and $n_i$ ($\mu_i$) is a quark number density (quark number chemical potential) for a flavor $i$.
$\chi_{ij}\equiv\frac{\partial n_i}{\partial \mu_j}$ is measured by a lattice QCD simulation \cite{Allton:2005gk} and is found to satisfy $\chi_{ij}\propto \delta_{ij}$ approximately.

\begin{figure}
\centering
\includegraphics[width=5cm, angle=-90, clip]{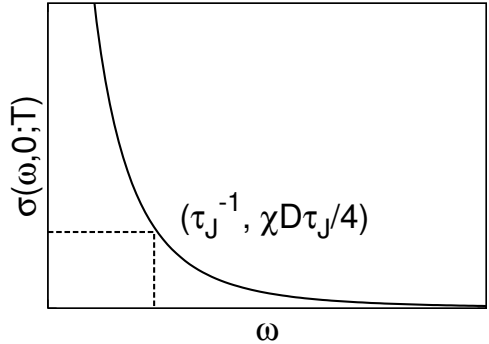}
\caption{
\footnotesize
Scaled transport-SPF $\sigma(\omega,\vec 0;T)$.
It diverges as $\sigma\propto 1/\omega$ in the limit $\omega\rightarrow 0$.
}
\label{DL:fig:transport-SPF}
\end{figure}

Let us consider the scaled spectral function $\sigma(q;T)\equiv -{\rm Im}G^{\mu\mu}_{\rm R}(q;T)/3q^2$ for the transport-SPF.
At $\vec k=\vec 0$,
\begin{eqnarray}
\sigma(\omega,\vec 0;T)=\frac{1}{2\omega}\frac{\chi D}{\tau^2_{\rm J}\omega^2+1},
\end{eqnarray}
and the scaled spectral function becomes larger for larger $\chi$ and $D$ and for smaller $\tau_{\rm J}$.
This can be physically understood as follows:
The number of photons and dileptons is large when there is strong electric current.
When $D$ is large and $\tau_{\rm J}$ is small, strong electric current is swiftly induced by the electric charge fluctuation, which is characterized by the susceptibility $\chi$.
As shown in Fig.~\ref{DL:fig:transport-SPF}, the scaled spectral function shows divergence $\sigma\propto\omega^{-1}$ at $\omega\sim 0$.

\subsection{Diffusion coefficient and relaxation time}
\label{DL:sec:3-b}

Our spectral function is parameterized by three unknown functions of temperature $\chi(T),\ D(T)$, and $\tau_{\rm J}(T)$.
The electric charge susceptibility $\chi$ can be expressed in terms of the quark number susceptibility by
\begin{eqnarray}
\chi=\frac{4}{9}\chi_{\rm uu}+\frac{1}{9}\chi_{\rm dd}+\frac{1}{9}\chi_{\rm ss}
-\frac{4}{9}\chi_{\rm ud}-\frac{4}{9}\chi_{\rm us}+\frac{2}{9}\chi_{\rm ds}
\approx \frac{2}{3}\chi_{\rm uu},
\end{eqnarray}
which can be fitted to the result of lattice QCD simulation \cite{Allton:2005gk} as follows
\begin{eqnarray}
\chi(T,\mu=0)\approx 0.28T^2\left[1+\tanh\left(\frac{T-T_{\rm c}^*}{0.15T_{\rm c}^*}\right)\right].
\end{eqnarray}
We take $T_{\rm c}^*$ to be the temperature at the center of the chiral crossover transition $T_{\rm c}^*\approx0.155$ GeV (see Fig.~4~(right) in \cite{Borsanyi:2010bp}).
We parameterize $D$ and $\tau_{\rm J}$ as 
\begin{eqnarray}
D\propto\frac{1}{T},
\tau_{\rm J}\propto\frac{1}{T},
\end{eqnarray}
for dimensional reason \cite{footnote2}.
To see the typical magnitude of the transport coefficients of weakly and strongly coupled plasmas, we refer to the results obtained by perturbative QCD \cite{Arnold:2000dr,Arnold:2003zc,Hong:2010at} with $g\approx 2$
\begin{eqnarray}
D^{\rm pQCD}&\approx& D^{\rm pQCD}_{\rm q}\approx \frac{0.150}{\alpha_{\rm s}^2\ln(0.461/\alpha_{\rm s})T}\approx \frac{4}{T},\\
\tau^{\rm pQCD}_{\rm J}&\approx& 3.748\times D^{\rm pQCD}\approx \frac{15}{T},
\end{eqnarray}
and those by AdS/CFT \cite{Natsuume:2007ty}
\begin{eqnarray}
D^{\rm AdS/CFT}=\frac{1}{2\pi T}, \ \tau^{\rm AdS/CFT}_{\rm J}=\frac{\ln 2}{2\pi T}.
\end{eqnarray}
Here $D^{\rm pQCD}_{\rm q}$ stands for {\it quark number} diffusion coefficient.
We can show $D=D_{\rm q}$ within the same approximation as used in the derivation of the constitutive equation in multi-flavor case. (See Appendix~\ref{DL:app:2}.)

\subsection{Comparison to experimental data}
\label{DL:sec:3-c}

The minimum bias theoretical spectrum is calculated by
\begin{eqnarray}
\label{DL:eq:minbias}
\frac{dN^{\rm min. bias}}{dm_{\rm ee}}&=&
\frac{1}{10}\times\left[
\frac{dN^{0-10\%}}{dm_{\rm ee}}+\cdots+
\frac{dN^{90-100\%}}{dm_{\rm ee}}
\right] \nonumber \\
&\approx&
\frac{1}{10}\times\left[
\frac{dN^{0-10\%}}{dm_{\rm ee}}+\cdots+
\frac{dN^{50-60\%}}{dm_{\rm ee}}
\right],
\end{eqnarray}
neglecting peripheral collisions with centrality 60-100 \% because their contributions are small.

\begin{figure}
\centering
\includegraphics[width=5cm, angle=-90, clip]{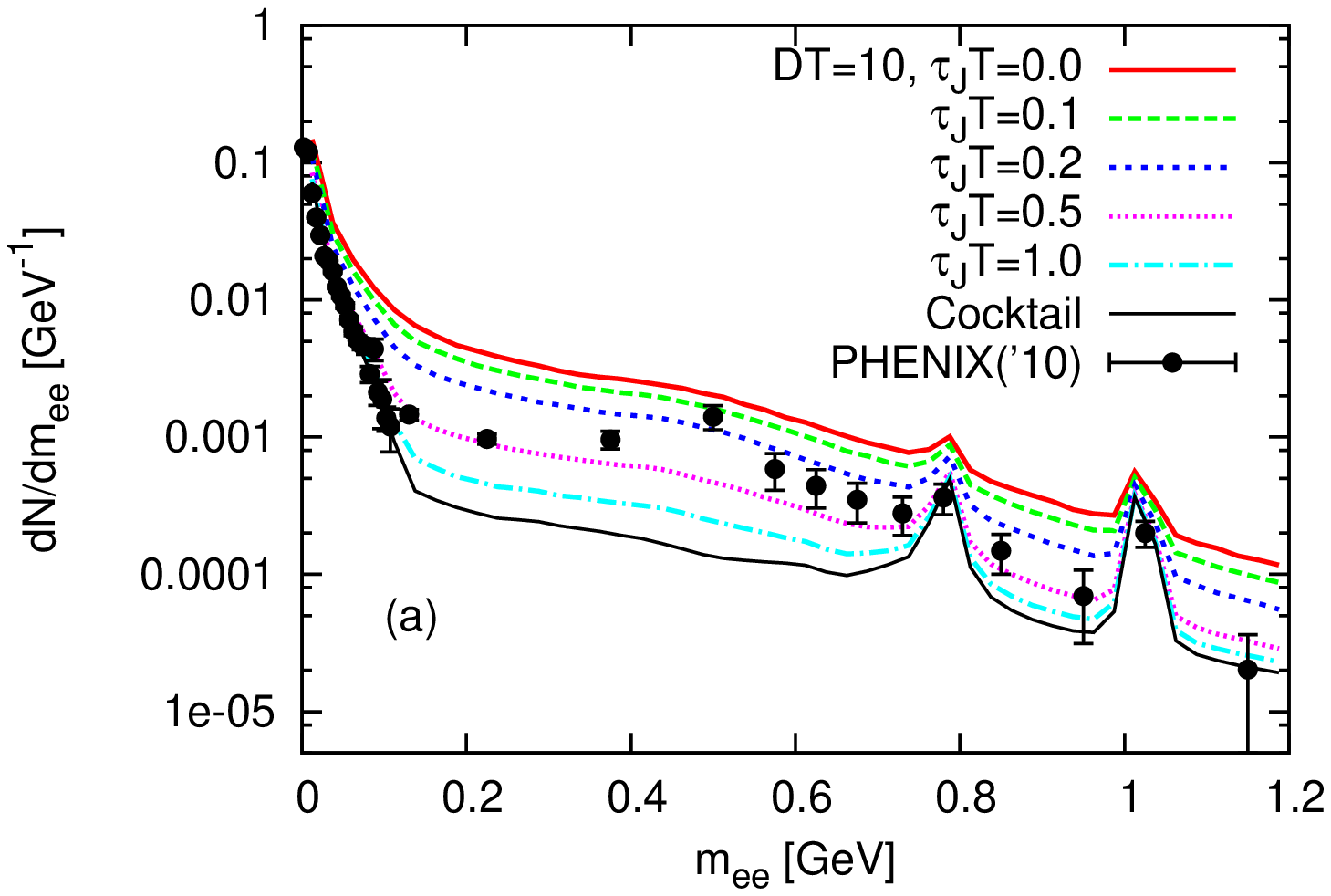}
\includegraphics[width=5cm, angle=-90, clip]{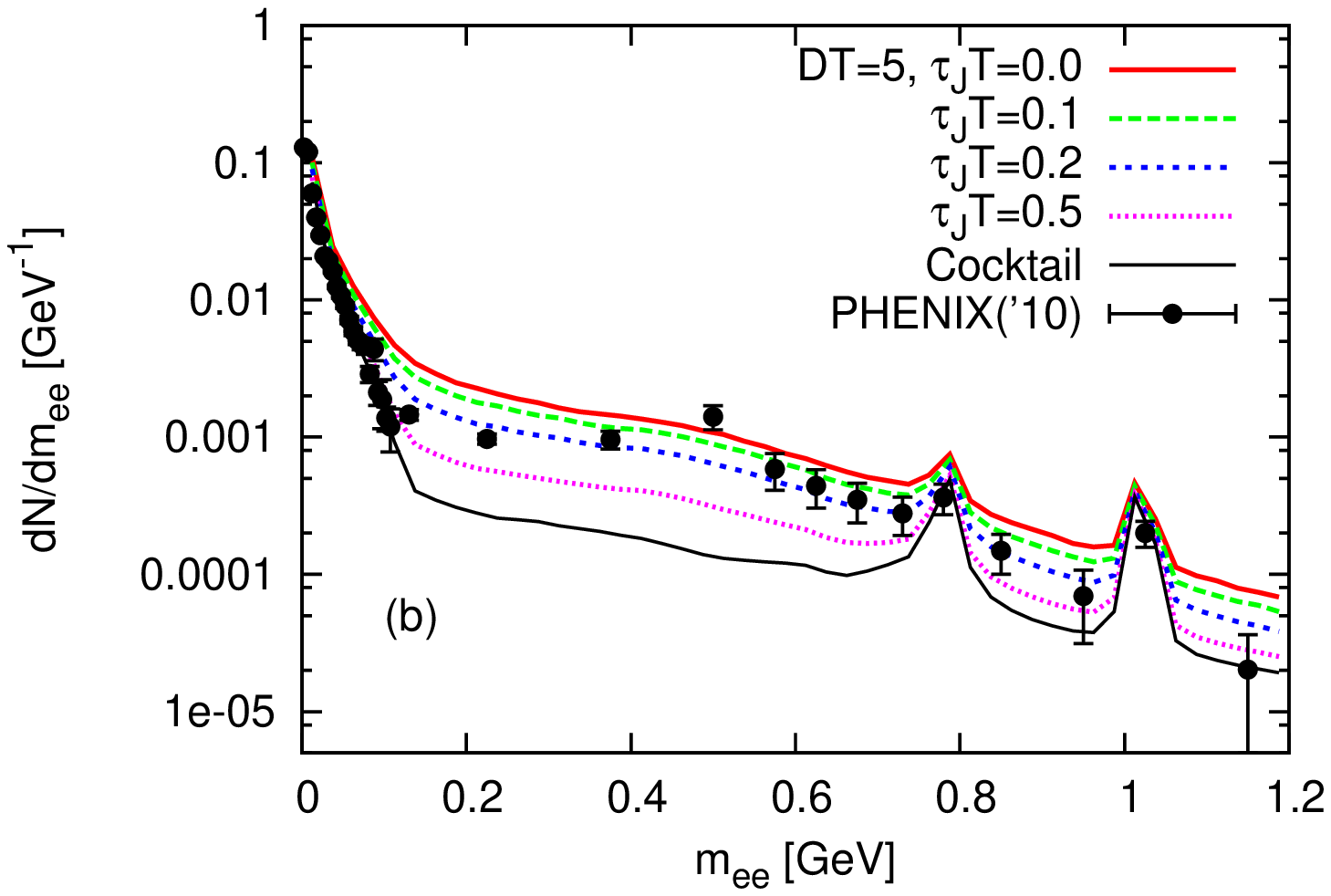}
\includegraphics[width=5cm, angle=-90, clip]{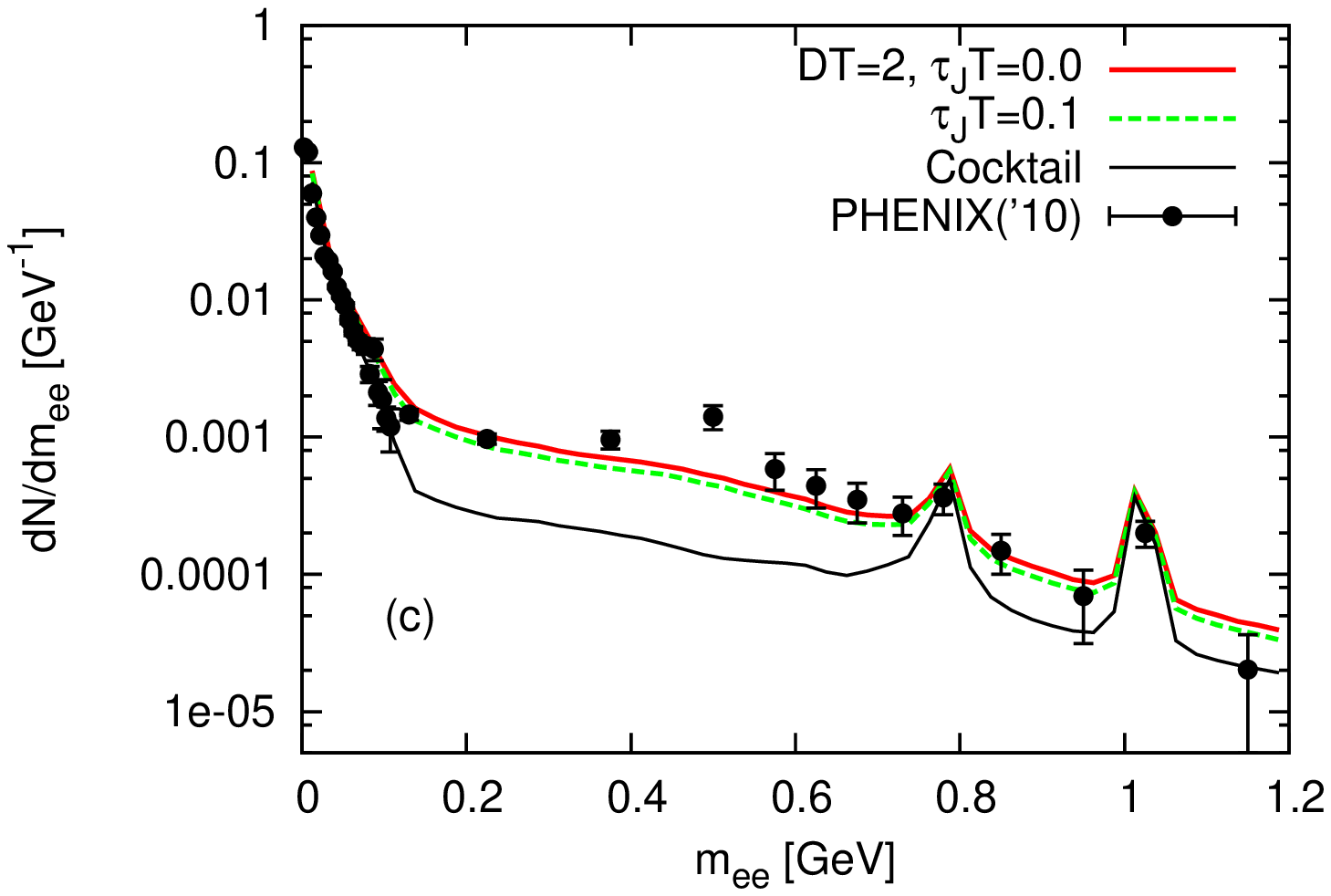}
\includegraphics[width=5cm, angle=-90, clip]{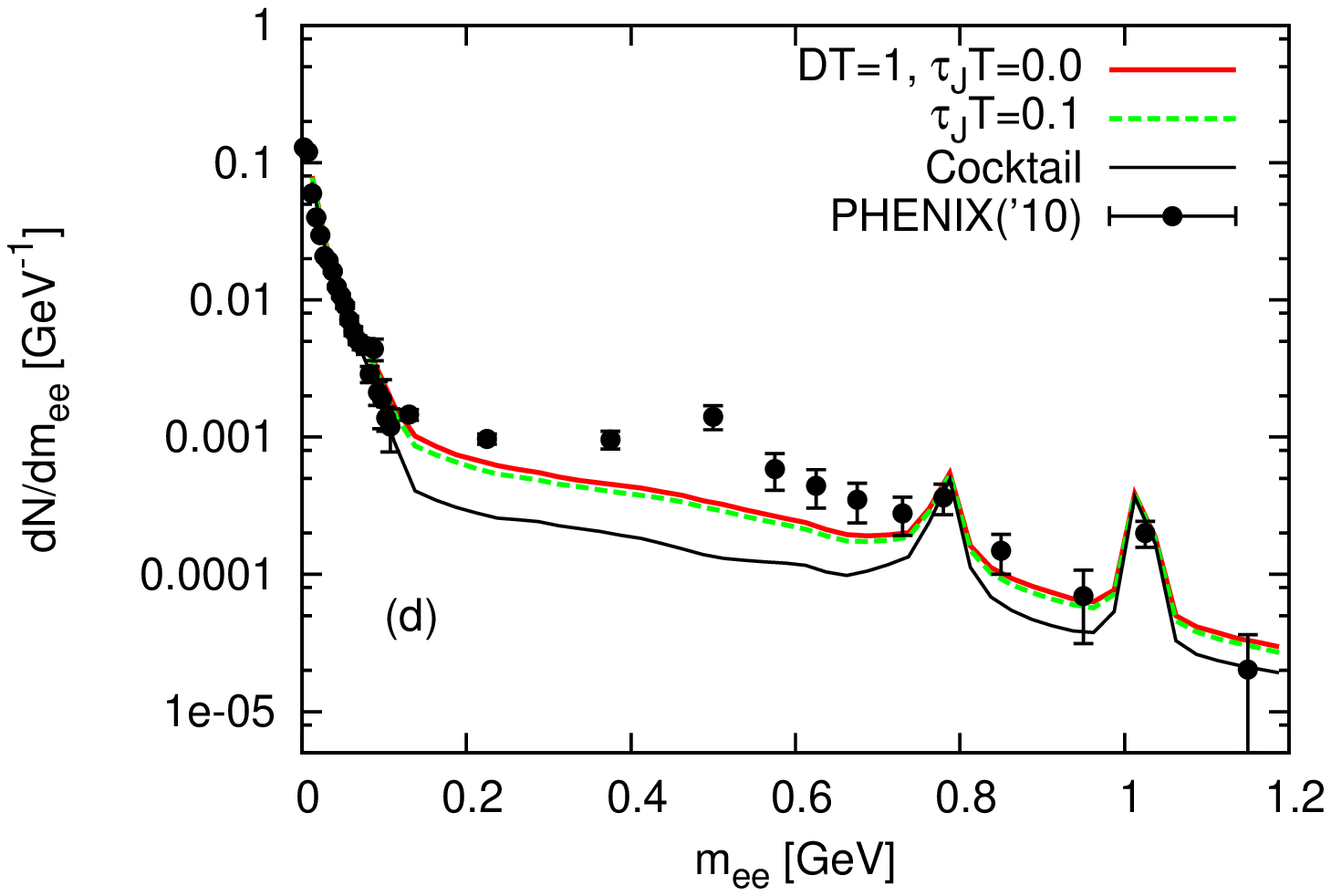}
\caption{
\footnotesize
(Color online)
We compare in (a)-(d) the dielectron spectrum $dN/dm_{\rm ee}$ calculated with the transport-SPF and the corresponding experimental data \cite{Adare:2009qk}.
In the transport-SPF, diffusion coefficient is fixed to (a) $D=10/T$, (b) $5/T$, (c) $2/T$, and (d) $1/T$.
The relaxation time $\tau_{\rm J}$ is varied in $(0.0$-$1.0)/T$.
We take into account the contributions from the hadronic decays after freezeout (denoted as ``Cocktail'' in these figures).
We only plot the statistical errors for experimental data.
}
\label{DL:fig:dNdM}
\end{figure}

In Fig.~\ref{DL:fig:dNdM}~(a)-(d), we compare the dielectron spectrum $dN/dm_{\rm ee}$ calculated with the transport-SPF with the experimental data.
The parameterization for the diffusion coefficient is (a) $D=10/T$, (b) $D=5/T$, 
(c) $D=2/T$, and (d) $D=1/T$ and that for the relaxation time is varied in $\tau_{\rm J}=(0.0$-$1.0)/T$.
We add the dielectrons from hadronic decays after freezeout (hadronic cocktail) \cite{Adare:2009qk} to the thermal contribution from the transport-SPF.
We find that our results with $(DT,\tau_{\rm J}T)=(10,0.5)$, $(5,0.2)$, and $(2,0.0$-$0.1)$ give dielectron spectra comparable to the experimental data.
These values are rather different from the theoretical estimates,  $(D^{\rm pQCD}T,\tau_{\rm J}^{\rm pQCD}T)\approx(4, 15)$ and $(D^{\rm AdS/CFT}T,\tau_{\rm J}^{\rm AdS/CFT}T)= (\frac{1}{2\pi}, \frac{\ln 2}{2\pi})$.
From (c) and (d), we also find that experimental data set lower bound for the diffusion coefficient $D\geq 2/T$ because the dielectron rate is largest with $\tau_{\rm J}=0$ for fixed $D$.

\begin{figure}
\centering
\includegraphics[width=5cm, angle=-90, clip]{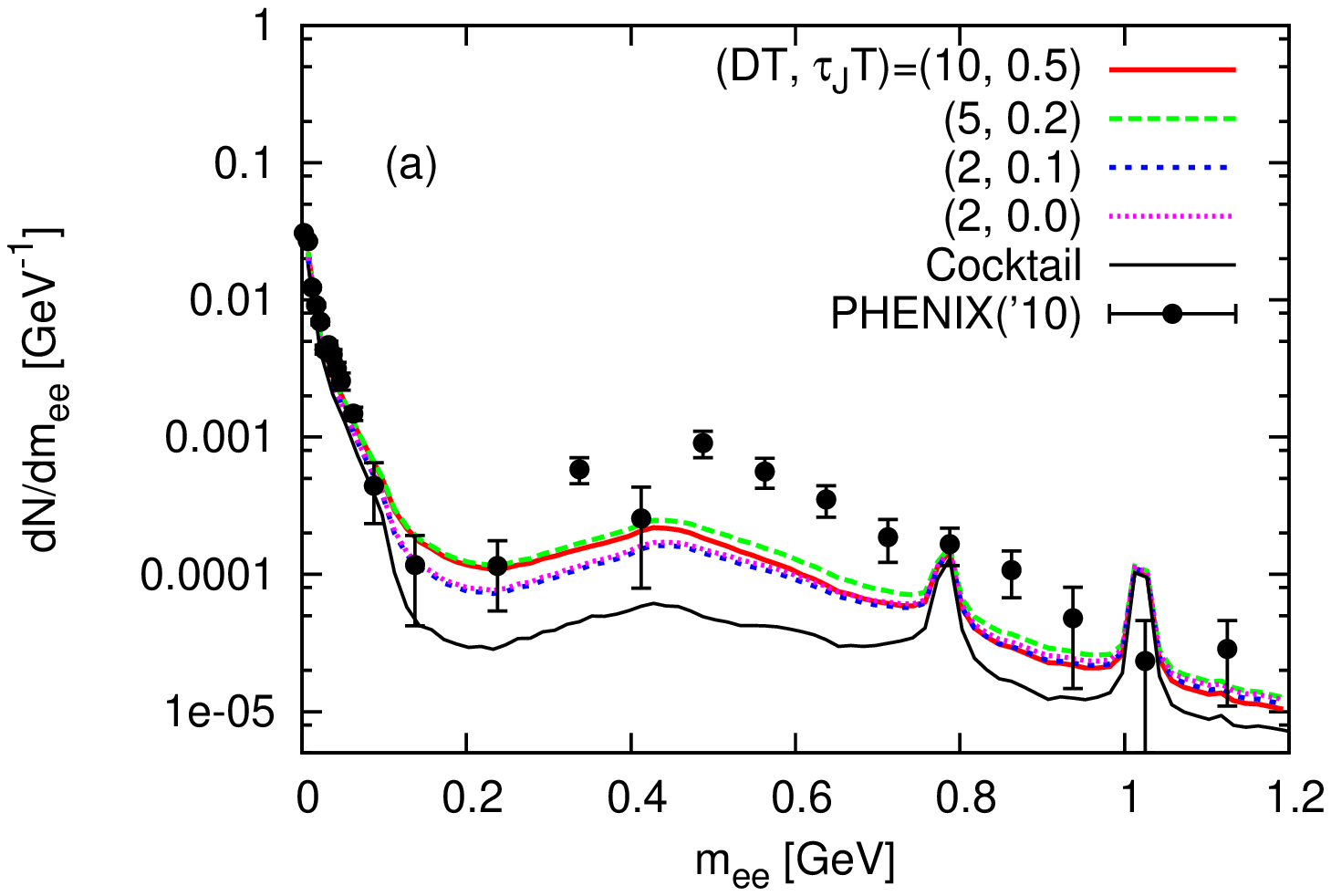}
\includegraphics[width=5cm, angle=-90, clip]{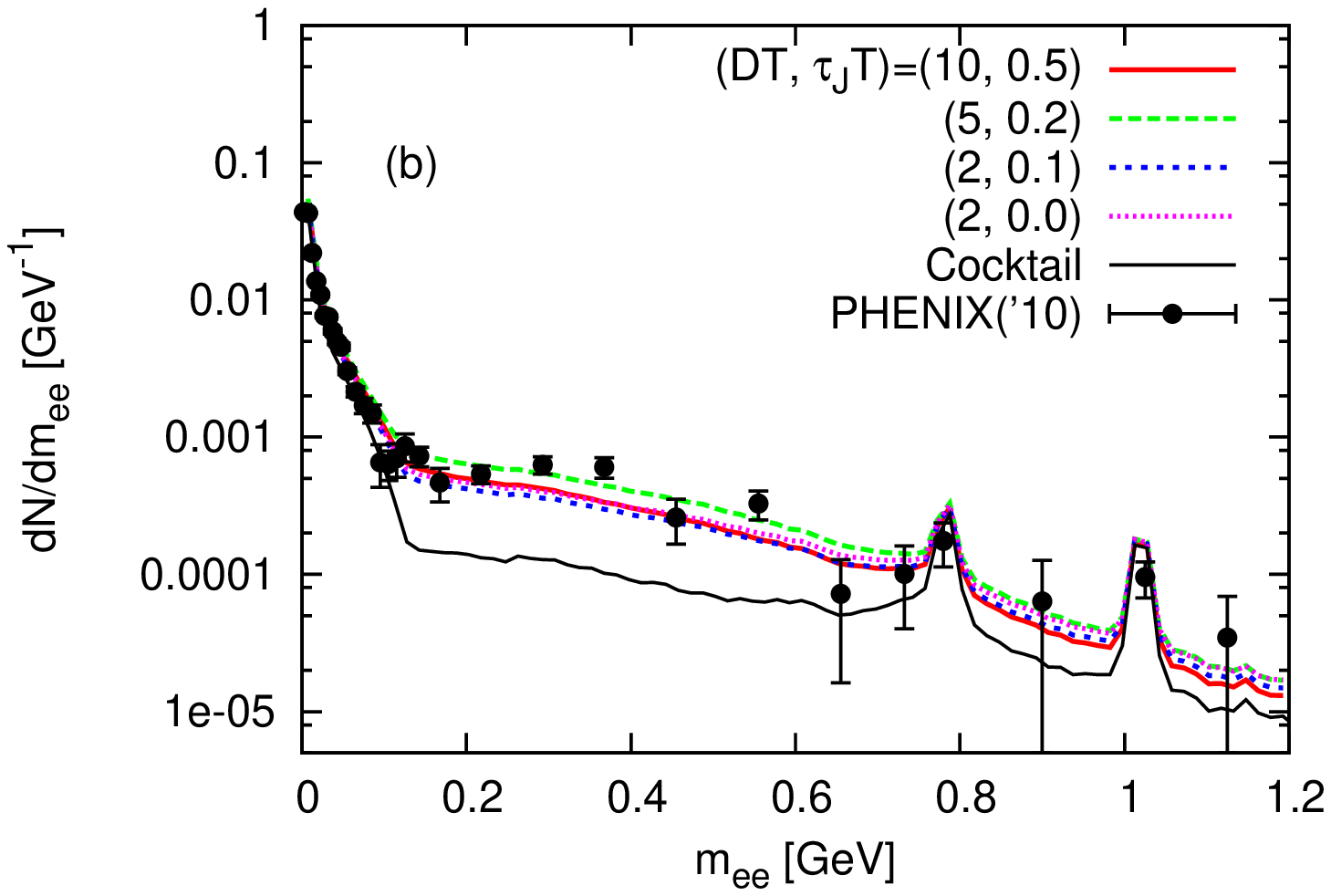}
\caption{
\footnotesize
(Color online)
In (a) and (b), we compare our calculation of the dielectron spectrum with restricted $p_{_{\rm T}}$-windows and the experimental data \cite{Adare:2009qk}.
The dielectron spectrum $dN/dm_{\rm ee}$ is restricted in (a) $0.0<p_{_{\rm T}}<0.5$ GeV and (b) $0.5<p_{_{\rm T}}<1.0$ GeV.
We calculate with parameter sets $(DT,\tau_{\rm J}T)=(10,0.5)$, $(5,0.2)$, and $(2,0.0$-$0.1)$ in the transport-SPF.
The hadronic decays after freezeout are taken into account (denoted as "Cocktail" in these figures).
In (a) and (b), we only plot the statistical errors for experimental data.
}
\label{DL:fig:pt0010}
\end{figure}

Next we study the dielectron spectrum in more detail.
In Fig.~\ref{DL:fig:pt0010}~(a) and (b), we show the dielectron invariant mass spectrum with the dielectron transverse momentum restricted in (a) $0\leq p_{_{\rm T}}\leq 0.5$ GeV and (b) $0.5\leq p_{_{\rm T}}\leq 1.0$ GeV.
We calculate these spectra with parameter sets $(DT,\tau_{\rm J}T)=(10,0.5)$, $(5,0.2)$, and $(2,0.0$-$0.1)$ in the transport-SPF.
Clearly all of our calculations for low $p_{_{\rm T}}$-window (a) undershoot the experimental data in $0.3<m_{\rm ee}<0.7$ GeV, while for high $p_{_{\rm T}}$-window (b) they all show good agreement with the data.

\begin{figure}
\centering
\includegraphics[width=5cm, angle=-90, clip]{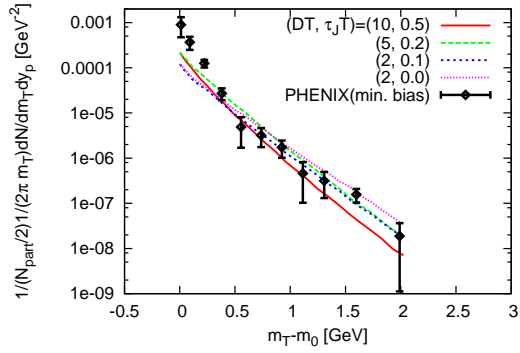}
\caption{
\footnotesize
(Color online)
We compare the transverse mass spectrum calculated with the transport-SPF and the corresponding experimental data \cite{Adare:2009qk}.
The parameter sets $(DT,\tau_{\rm J}T)$ are the same with those in Fig.~\ref{DL:fig:pt0010}
In the experimental data, other sources such as hadron decays are subtracted and PHENIX acceptance is corrected.
We only plot the statistical errors for experimental data.
}
\label{DL:fig:dNdMT}
\end{figure}

In Fig.~\ref{DL:fig:dNdMT}, we show the dielectron transverse mass ($m_{_{\rm T}}$) distribution at mid-rapidity $y_{\rm p}=0$.
The calculation is performed with $(DT,\tau_{\rm J}T)=(10,0.5)$, $(5,0.2)$, and $(2,0.0$-$0.1)$.
We first calculate the dielectron $p_{_{\rm T}}$-spectrum by integrating in the invariant mass range $0.3\leq m_{\rm ee}\leq 0.75$ GeV for each centrality.
We then divide the dielectron $p_{_{\rm T}}$-spectra by half of the number of participant nucleons ($N_{\rm part}$/2) for each centrality.
We take the horizontal axis to be $m_{_{\rm T}}-m_0=\sqrt{p^2_{_{\rm T}}+m_0^2}-m_0$ to obtain $m_{_{\rm T}}$-spectra.
Here $m_0$ is the mean value of the dielectron invariant mass with minimum bias $m_0\equiv\langle m_{\rm ee}\rangle$ in $0.3\leq m_{\rm ee}\leq 0.75$ GeV, which turns out to be $m_0\approx 0.47$ GeV in our calculation.
Experimental data are also obtained by the same procedure after subtracting other sources such as hadron decays and correcting PHENIX acceptance.
The experimental data exceed the theoretical spectra at $m_{_{\rm T}}-m_0<0.3$ GeV while the former are comparable with the latter at $m_{_{\rm T}}-m_0>0.3$ GeV.
This is consistent with what we found in Fig.~\ref{DL:fig:pt0010}.

\begin{figure}
\centering
\includegraphics[width=5cm, angle=-90, clip]{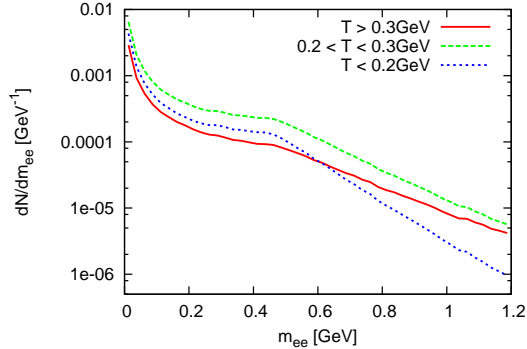}
\caption{
\footnotesize
(Color online)
Plotted is dielectron production from the matter with $T>0.3$ GeV, $0.2<T<0.3$ GeV, and $T<0.2$ GeV.
The parameter for transport-SPF is $(DT,\tau_{\rm J}T)=(10,0.5)$.
The PHENIX acceptance is taken into account.
}
\label{DL:fig:tdep}
\end{figure}

In Fig.~\ref{DL:fig:tdep}, we show dielectron production from the hot matter with different temperature range: $T>0.3$ GeV, $0.2<T<0.3$ GeV, and $T<0.2$ GeV.
Clearly the intermediate-temperature QGP phase with $0.2<T<0.3$ GeV is the main source for the dielectron production in spite of its small spacetime volume.
This makes a sharp contrast with the expectation that the low-mass dielectrons are radiated from the low-temperature hadronic phase due to the larger spacetime volume \cite{Rapp:2010sj}.
Furthermore, this tendency may be consistent with the previous analyses at SPS, where the transport peak does not play an important role.

%%%%%%%%%%%%%%%%%%%%%%%%%%%%%%%%%%%%%%%%%%%%%%%%%%%%%%%%%%
\section{Conclusion and Outlook}
\label{DL:sec:4}

\begin{figure}
\centering
\includegraphics[width=5cm, angle=-90, clip]{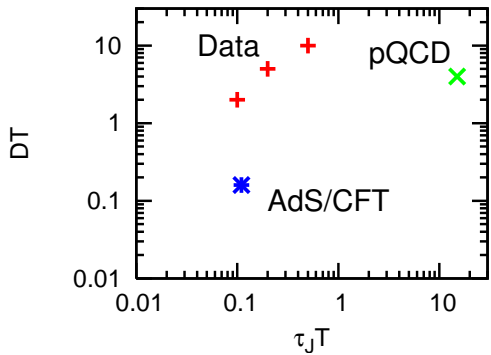}
\caption{
\footnotesize
(Color online)
We plot parameter sets $(DT,\tau_{\rm J}T)$ with which the experimental data $dM/dm_{\rm ee}$ are consistent.
We also plot two limiting sets obtained by weak-coupling (pQCD) and strong-coupling (AdS/CFT) calculation.
}
\label{DL:fig:range}
\end{figure}

In this paper, we have studied the dilepton production from the hot dynamical medium in relativistic heavy ion collisions.
In particular, we have explored the novel relation between the low-mass dielectron enhancement and the transport property of the hot QCD matter.
Because the dielectron rate from the transport-SPF is divergent at $\omega\sim 0$, it is reasonable to expect that the transport property is relevant to the low-mass dielectron enhancement.
Also the transport theoretical approach enables us to treat the problem non-perturbatively.

We have conducted our analysis in the following way.
We start from the second-order formalism of relativistic dissipative hydrodynamics, which introduces diffusion coefficient $D$ and relaxation time $\tau_{\rm J}$.
By performing linear analysis, we parameterized the low-frequency and long-wavelength region of the spectral function with these transport coefficients.
We calculated the dielectron yield by combining the transport-SPF and the full (3+1)-dimensional hydrodynamic medium evolution with the lattice EoS and attempted to extract the transport coefficients from the data.
We summarize our results in Fig.~\ref{DL:fig:range}.
We found that transport peak with parameter sets $(DT,\tau_{\rm J}T)=(10,0.5)$, $(5,0.2)$, $(2,0.0$-$0.1)$ can reproduce the experimental data $dN/dm_{\rm ee}$.
These obtained parameter sets are rather different from both weak-coupling pQCD calculation and strong-coupling calculation on the basis of the AdS/CFT correspondence.
We also showed that in order to explain the data, the diffusion coefficient must be $D\geq 2/T$.
While these parameter sets could reproduce the more detailed experimental data $dN/dm_{\rm ee}$ restricted in $0.5<p_{_{\rm T}}<1.0$ GeV and $(1/2\pi m_{_{\rm T}})dN/dm_{_{\rm T}}dy_{\rm p}$ in $m_{_{\rm T}}-m_0>0.3$ GeV, they could not reproduce those restricted in $0.0<p_{_{\rm T}}<0.5$ GeV and in $m_{_{\rm T}}-m_0<0.3$ GeV.
In view of these calculations, we conclude that the low-mass dielectrons at RHIC has not been fully understood theoretically despite the fact that the transport peak of the spectral function has a right tendency to enhance low-mass dileptons.
   
Meanwhile we also found that a large portion of the thermal dielectron radiation comes from the high-temperature QGP phase with $T>0.2$ GeV because the fluctuation of electric charge is larger at higher temperature.
Note that RHIC has explored the temperature range $T>0.2$ GeV for the first time ever.
This tendency of dielectron production from the transport peak may be consistent with the dielectron production at SPS, which can be explained without the transport peak.

We list several future implications of interest:
(i) First-principle calculation of transport coefficients and spectral function by lattice QCD is highly desirable.
Calculation of the spectral function with finite momentum may give a clue to understand the origin of low-mass dielectron enhancement.
Some recent developments in such a direction can be seen in Refs. \cite{Aarts:2007wj,Aarts:2010mr,Ding:2010ga,Ding:2011hr}.
(ii) The event-by-event fluctuation of the initial geometry, which is expected to remain large at the early QGP phase, would enhance the dielectron production from the transport peak, as it does for real photon emission from high temperature region \cite{Chatterjee:2011dw}.
(iii) Since high-temperature QGP phase emits a large portion of thermal dielectrons in spite of its small spacetime volume, the dielectron production from pre-equilibrium stage, or the so-called glasma, might also be important and thus needs to be evaluated.

\section*{Acknowledgment} 
A part of the numerical computation in this work was carried out at the
Kobayashi-Maskawa Institute Computer Facility and at the Yukawa Institute Computer Facility.
Y.~Akamatsu thanks G.~Aarts for suggesting that the transport property is encoded in the spectral function.
He also thanks Y.~Akiba for the fruitful discussion that led to re-examination of the low-mass dielectron production.
Y.~Akamatsu is partially supported 
by JSPS fellowships for Young Scientists.
T.~Hatsuda is partially supported 
by No.~2004, Grant-in-Aid for Scientific Research on Innovative Areas.
T.~Hirano is partially supported 
by Sumitomo Foundation No.~080734,
by Grant-in-Aid for Scientific Research No.~22740151, and 
by JSPS Excellent Young Researchers Overseas Visit Program.
T.~Hirano also thanks the Nuclear Theory Group at LBNL for its hospitality.

\appendix

%%%%%%%%%%%%%%%%%%%%%%%%%%%%%%%%%%%%%%%%%%%%%%%%%%%%%%%%%%
\section{Model of low-mass dileptons: vector meson dominance}
\label{DL:app:1}

Here we assume vector meson dominance on the hadronic coupling to virtual photon and study its consequence on the dilepton spectra.
Two standard scenarios, the dropping mass and the width broadening of the vector meson spectral function (VMD-SPF) will be considered.
 
\subsection{Vector meson at finite temperature}
\label{DL:app:1-a} 

We define the dimensionless spectral function at finite $T$ as
$\sigma(q;T)\equiv-{\rm Im}G_{\rm R}^{\mu \mu}(q;T)/3q^2$.
In the vacuum  at  $T=0$, it can be is measured by the cross section 
of the electron-positron annihilation ($e^+e^-\rightarrow$ hadrons) and can be decomposed as
\begin{eqnarray}
\sigma(q;T=0)=\sigma_{\rho}(q;T=0)+\frac{1}{9}\sigma_{\omega}(q;T=0)+\frac{1}{9}\sigma_{\sigma}(q;T=0),
\end{eqnarray}
where $\sigma_{\rho}(q;T)$, $\sigma_{\omega}(q;T)$, $\sigma_{\phi}(q;T)$
and $\sigma(q;T)$ correspond to the current correlation of  $J_{\rho}^{\mu}=\frac{1}{2}(\bar {\rm u}\gamma^{\mu}{\rm u}-
\bar {\rm d}\gamma^{\mu}{\rm d})$, 
$J_{\omega}^{\mu} =\frac{1}{2}(\bar {\rm u}\gamma^{\mu}{\rm u}+
\bar {\rm d}\gamma^{\mu}{\rm d})$,
$J_{\phi}^{\mu} = \bar {\rm s}\gamma^{\mu}{\rm s}$ and 
 $J^{\mu}=J_{\rho}^{\mu}+\frac{1}{3}J_{\omega}^{\mu}-\frac{1}{3}J_{\phi}^{\mu}$, respectively.
 
The spectral function at zero temperature, $\sigma(q;T=0)$, has hadronic contribution at low $q$ and perturbative QCD continuum at high $q$ with an approximate continuum threshold at $ 1-1.5$ GeV and may be 
parametrized as \cite{Shuryak:1993kg}
\begin{eqnarray}
\label{DL:eq:SPF_VMD}
\sigma_{_V}(q;T=0)
=\left\{
\begin{array}{l}
\frac{f_{_V}^2\Gamma_{_V}m_{_V}}{(q^2-m_{_V}^2)^2+\Gamma_{_V}^2m_{_V}^2}
\ \ \ (s_{_{0V}}  < q^2 < s_{_{1V}})\\
c_{_V} \ \ \ \ \ \ \ \ \ \ \ \ \ \ \ \ \ \ \ (s_{_{1V}} < q^2)
\end{array}
\right. ,
\end{eqnarray} 
with $V=\rho, \omega, \phi$.
Here  $s_{_{0V}}$ corresponds to the threshold associated with the decay into two pions (280 MeV) or three pions (420 MeV), while $s_{_{1V}}$ corresponds to the continuum threshold.
We take the height of the continuum $c_V$ from the leading order perturbation theory and fit the other parameters (the resonance mass $m_{_V}$, the resonance width $\Gamma_{_V}$, the resonance height $f_{_V}$ and the continuum threshold $s_{_{1V}}$)
by using  experimental data in \cite{Nakamura:2010}.
These parameters are listed in the Table~\ref{DL:table:SPF_VMD}.

Since the $\rho$ channel is a dominant source of the thermal component of the low-mass dileptons and is expected to receive the medium effect most strongly, we hereafter restrict our analysis on the medium modifications of $\sigma_{\rho}(q;T)$ at finite $T$ within the same parametrization as Eq.~(\ref{DL:eq:SPF_VMD}).
Then the simplified form of the dilepton production rate reads
\begin{eqnarray}
\label{DL:eq:formula_T=L}
\frac{E_1E_2dR_{l^+l^-}}{d^3p_1d^3p_2}
&=&-\frac{\alpha^2}{6\pi^4q^2}\left[
\left(1+\frac{2m_l^2}{q^2}\right){\rm Im}G_{\rm R}^{\mu\mu}(q;T)\right]f_{\rm BE}(q^0;T)\nonumber \\
&\simeq &\frac{\alpha^2}{2\pi^4}\left[
\left(1+\frac{2m_l^2}{q^2}\right)\sigma_{\rho}(q;T)\right]f_{\rm BE}(q^0;T),
\end{eqnarray}
where we implicitly assume that the transverse and longitudinal spectral functions are the same even at finite $T$.

\begin{table}
\caption{
\footnotesize
The parameters of the vacuum spectral functions given in the unit of GeV except for dimensionless parameters $c_{_V}$.
They are obtained by fitting the experimental data \cite{Nakamura:2010}.
}
\ \\
\centering
\begin{tabular}{c|c|c|c|c|c|c}
\hline\hline
vector meson & \ $m_{_V}$ \ & \ $\Gamma_{_V}$ \ & \ $f_{_V}$ \ & \ $\sqrt{s_{_{0V}}}$ \ 
& \ $\sqrt{s_{_{1V}}}$ \ & \ $c_{_V}$ \ \\ \hline 
$\rho$ & 0.77 & 0.15 & 0.15 &  0.28 & 1.3 & $1/8\pi$ \\
$\omega$ & 0.78 & 0.008 & 0.14 & 0.42 & 1.1 & $1/8\pi$ \\ 
$\phi$ & 1.02 & 0.004 & 0.24 & 0.42 & 1.5 & $1/4\pi$ \\
\hline
\end{tabular} 
\label{DL:table:SPF_VMD}
\end{table}

Let us now consider  so-called dropping mass scenario where the mass parameters, $m_{\rho}(T)$ and $s_{_{1\rho}}(T)$, are assumed to scale with the chiral condensate \cite{Pisarski:1981mq, Brown:1991kk, Hatsuda:1991ez, Marco:1999xz}:
\begin{eqnarray}
\label{DL:eq:dropping_mass}
\frac{m_{\rho}(T)}{m_{\rho}(0)}
=\frac{s_{_{1\rho}}(T)}{s_{_{1\rho}}(0)}
=\frac{\langle\bar qq\rangle_{_T}}{\langle\bar qq\rangle_0}. 
\end{eqnarray}

The scaling function in Eq.~(\ref{DL:eq:dropping_mass}) is obtained by fitting the latest lattice data (Fig.~4~(right) in \cite{Borsanyi:2010bp}) by the following ansatz
with $(T_{\rm c}^*,\Delta)=(0.155 \ {\rm GeV}, 0.025 \ {\rm GeV})$:
\begin{eqnarray}
\label{eq:chiral_condensate}
\frac{\langle\bar qq\rangle_T}{\langle\bar qq\rangle_0}
=\frac{1}{2}\left[1-{\rm tanh}\left(\frac{T-T_{\rm c}^*}{\Delta}\right)\right] .
\end{eqnarray}

\begin{figure}
\centering
\includegraphics[width=5cm, angle=-90, clip]{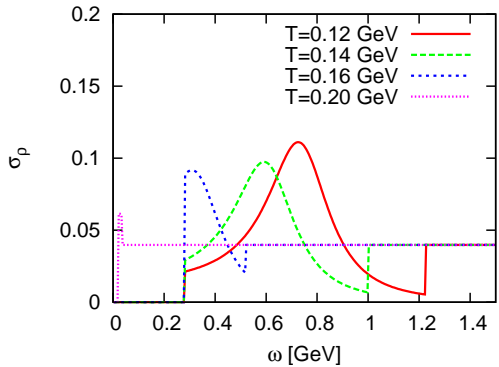}
\caption{
\footnotesize
(Color online)
Shown is the scaled spectral function $\sigma_{\rho}(\omega,\vec 0;T)$ as a function of $\omega$ at several different temperatures.
The parameters in the spectral function are $(T_{\rm c}^*, \Delta,\Gamma_{\rho})=(0.155 \ {\rm GeV}, 0.025 \ {\rm GeV}, 0.3 \ {\rm GeV})$.
}
\label{DL:fig:VMD-SPF}
\end{figure}

We assume that the pion mass does not change appreciably below $T_{\rm c}^*$, and adopt the following prescription for the low-energy threshold,
$ s_{_{0\rho}}={\rm min.}\{ 2m_{\pi}, \ m_{\rho}(T) \}$.
Instead of introducing a parametrization for the $T$-dependence of $\Gamma_{\rho}(T)$, we vary its value in the range $0.15$-$0.45$ GeV to see its effect on the dilepton yield.
Once it is given, remaining parameter, $f_{\rho}(T)$, is constrained by the QCD spectral sum rule \cite{Hatsuda:1992bv}:
\begin{eqnarray}
\label{DL:eq:sum_rule}
\int_0^{\infty}d\omega^2 \bigl[ \sigma_{\rho}(\omega,\vec 0;T)- c_{\rho}\bigr]=0.
\end{eqnarray}
Shown in Fig.~\ref{DL:fig:VMD-SPF} is a resultant  spectral function in the $\rho$ channel with a parameter set, $(T_{\rm c}^*,\Delta,\Gamma_{\rho})=(0.155 \ {\rm GeV}, 0.025 \ {\rm GeV}, 0.3 \ {\rm GeV})$.

\subsection{Dielectron spectra}
\label{DL:app:1-b}

\begin{figure}
\centering
\includegraphics[width=5cm, angle=-90, clip]{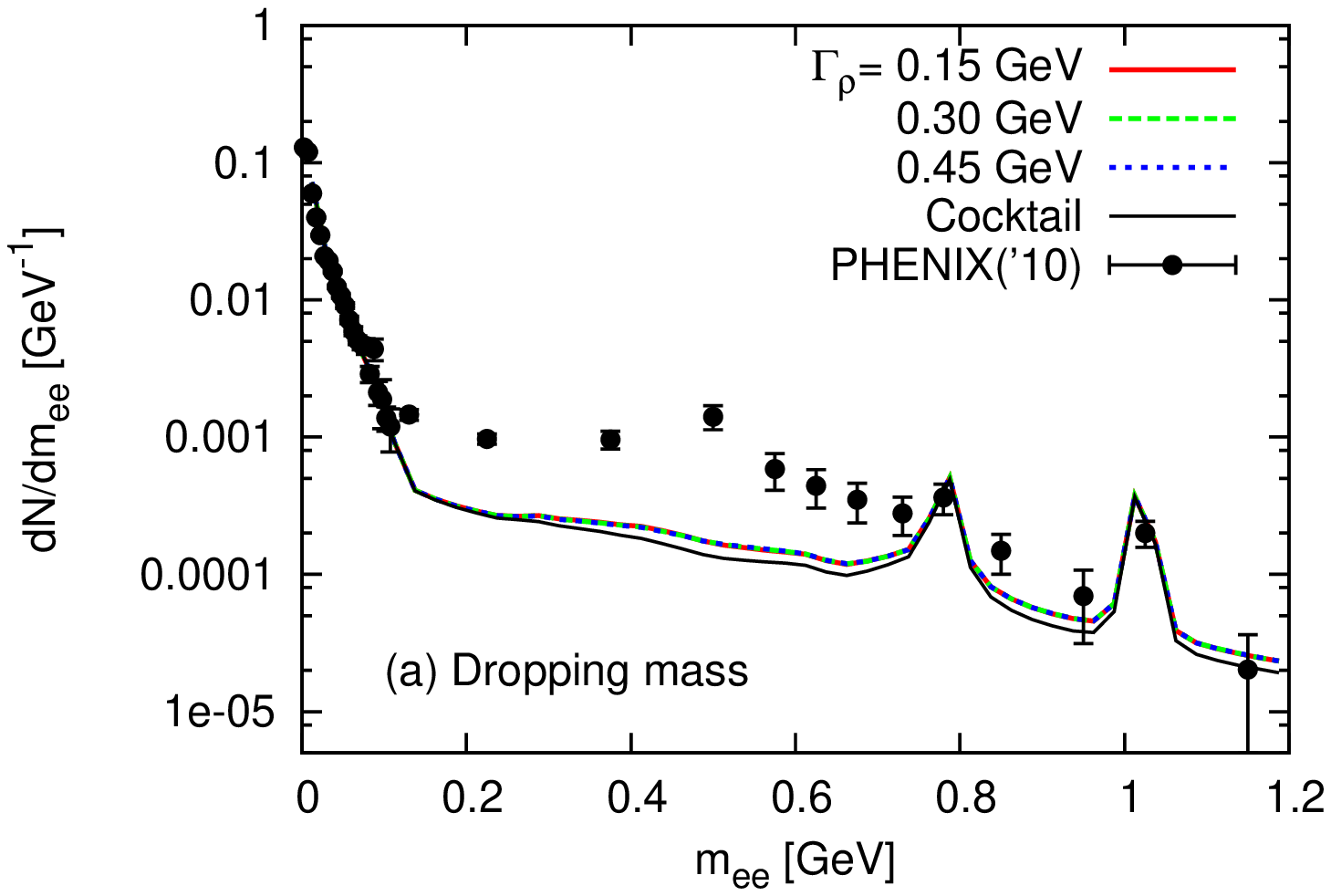}
\includegraphics[width=5cm, angle=-90, clip]{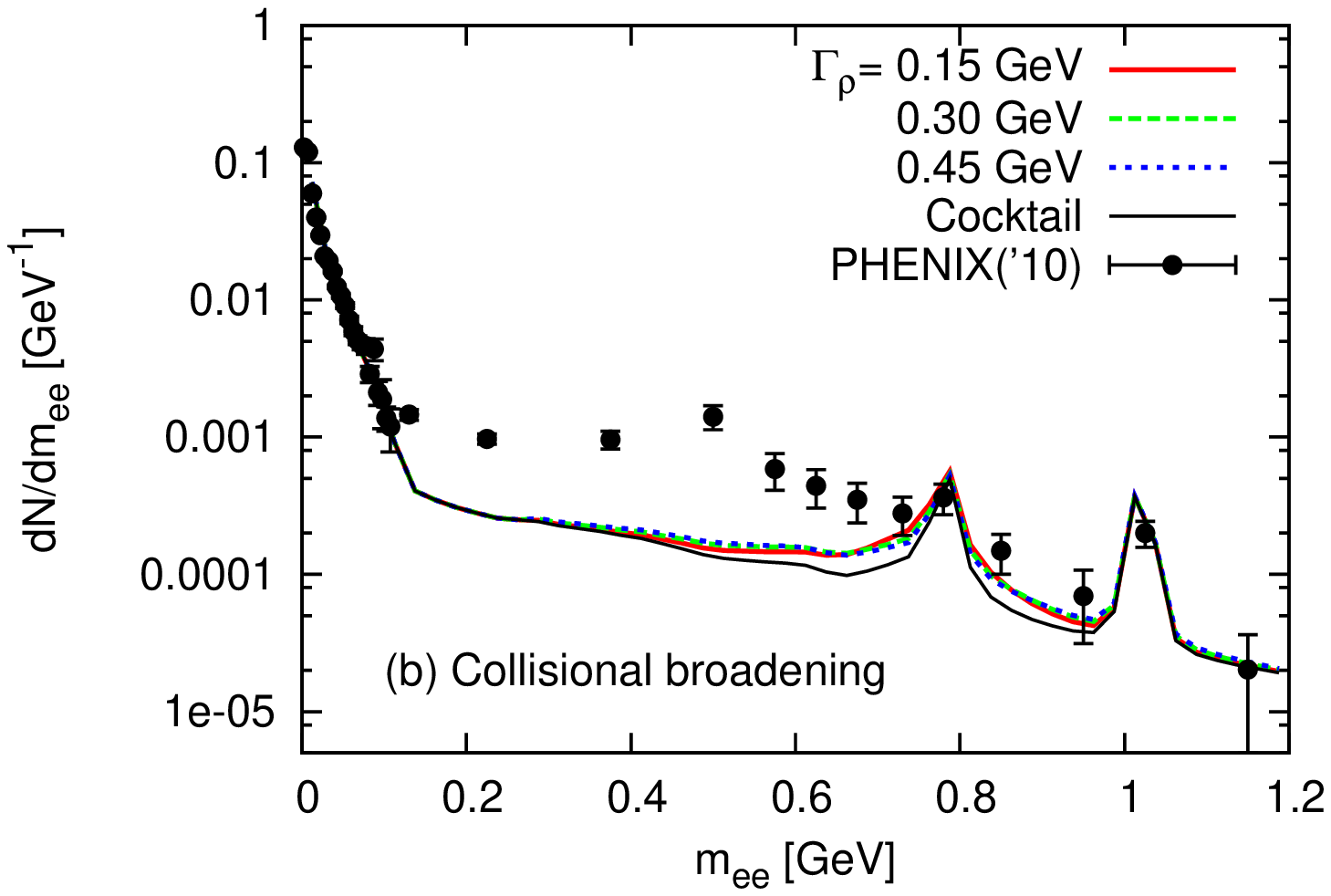}
\caption{
\footnotesize
(Color online)
Shown in (a) and (b) are sum of the dielectron spectra from the thermal medium and the contributions from the hadronic decays after freezeout.
The latter is denoted as ``Cocktail'' in these figures.
The parameters of the VMD-SPF are $\Gamma_{\rho}=0.15, 0.3, 0.45$ GeV and (a) $m_{\rho}(T)\propto\langle \bar qq\rangle_T$ (dropping mass) and (b) $m_{\rho}(T)=0.77$ GeV (collisional broadening).
These spectra are compared with experimental data of dielectron spectrum with minimum bias \cite{Adare:2009qk}.
In (a) and (b), we only plot the statistical errors for experimental data.
}
\label{DL:fig:dNdM_VMD}
\end{figure}

In Fig.~\ref{DL:fig:dNdM_VMD}, we compare the theoretical spectrum with the experimental data by taking into account the contributions from hadronic decays after freezeout (hadronic cocktail) given in \cite{Adare:2009qk}.
The PHENIX acceptance is already taken into account.
We show in Fig.~\ref{DL:fig:dNdM_VMD}~(a) the thermal dielectron spectrum for $\Gamma_{\rho}=0.15, 0.3, 0.45$ GeV and $m_{\rho}(T)=0.77$ GeV (collisional broadening).
It is clear from this figure that the dielectron emission with collisional broadening only cannot explain the experimental data in the low invariant mass region $m_{\rm ee}<0.6$ GeV.
We show in Fig.~\ref{DL:fig:dNdM_VMD}~(b) the thermal dielectron spectrum for $\Gamma_{\rho}=0.15, 0.3, 0.45$ GeV and $m_{\rho}(T)\propto\langle \bar qq\rangle_T$.
Even with the dropping mass, the dilepton yield undershoots the experimental data substantially in the low invariant mass region $m_{\rm ee}<0.6$ GeV.
The above findings are consistent with the previous attempts to reproduce the low-mass dileptons at PHENIX \cite{Rapp:2010sj,Ghosh:2010wt,Dusling:2007su,Dusling:2009ej,Bratkovskaya:2008bf,Bratkovskaya:2010gh}.

\section{Extension of Transport Spectral Function to Multi-Flavor Case}
\label{DL:app:2}
We derive the transport spectral function for multi-flavor case.
In this derivation, it is more convenient to treat flavor current for each quark species $N_f^{\mu}$.
We start from conservation laws in external field $\delta A^{\mu}$ which couples to each flavor current and gives perturbed Hamiltonian $\delta H(t)=\int d^3x \sum_{f}q_fN^{\mu}(x)\delta A_{\mu}(x)$:
\begin{eqnarray}
\partial_{\nu}T^{\nu\mu}&=&F^{\mu\nu}\sum_fq_fN_{f,\nu},\\
\partial_{\mu}N_f^{\mu}&=&0 \ \ \  (f={\rm u,d,s}),
\end{eqnarray}
with $F_{\mu\nu}\equiv \partial_{\mu}\delta A_{\nu}-\partial_{\nu}\delta A_{\mu}$ and $q_{\rm u}=2/3, q_{\rm d,s}=-1/3$.
According to relativistic viscous hydrodynamics in Landau frame \cite{Landau}, we decompose $T^{\mu\nu}$ and $N_f^{\mu}$ as
\begin{eqnarray}
\label{mSPF:eq:continuity_em}
T^{\mu\nu}&=&eu^{\mu}u^{\nu}-(P+\Pi)\triangle^{\mu\nu}+\pi^{\mu\nu},\\
\label{mSPF:eq:continuity_flavor}
N_f^{\mu}&=&n_fu^{\mu}+\nu_f^{\mu},\\
\triangle^{\mu\nu}&\equiv& g^{\mu\nu}-u^{\mu}u^{\nu},
\end{eqnarray}
with bulk pressure $\Pi$, shear stress tensor $\pi^{\mu\nu}$, and dissipative flavor current $\nu_f^{\mu}$ satisfying $\pi^{\mu\nu}u_{\nu}=0,\pi^{\mu}_{\mu}=0,\nu_f^{\mu}u_{\mu}=0$.
The entropy current in the second-order formalism \cite{Israel:1976tn,Israel:1979wp,Natsuume:2007ty} is decomposed as
\begin{eqnarray}
s^{\mu}&=&su^{\mu}-\sum_f\frac{\mu_f}{T}\nu_f^{\mu}\nonumber \\
&&-\frac{u^{\mu}}{2T}\left(\beta_0\Pi^2 -\sum_{f,f'}\beta^{ff'}_1\nu_f^{\mu}\nu_{f',\mu}+\beta_2\pi^{\rho\sigma}\pi_{\rho\sigma}\right)
-\frac{1}{T}\left(\sum_f\alpha^f_{0}\Pi\nu_f^{\mu}+\sum_f\alpha^f_{1}\pi^{\mu\nu}\nu_{f,\nu}\right),
\end{eqnarray}
with coupling coefficients $\alpha^f_{0},\alpha^f_{1}$ and $\beta_{0,2} \ (\geq 0)$, a positive semi-definite matrix $\beta^{ff'}_1$, and a chemical potential $\mu_f$ for each flavor.
Divergence of the entropy current by using Eqs.~(\ref{mSPF:eq:continuity_em}) and (\ref{mSPF:eq:continuity_em}) up to second order deviation from equilibrium is
\begin{eqnarray}
\partial_{\mu}s^{\mu}&=&
-\frac{\Pi}{T}\left(\triangle^{\mu\nu}\partial_{\mu}u_{\nu}+\beta_0\dot\Pi+\sum_f\alpha^f_{0}\partial_{\mu}\nu_f^{\mu}\right)
+\frac{\pi^{\mu\nu}}{T}\left(\partial_{\mu}u_{\nu}-\beta_2\dot\pi_{\mu\nu}-\sum_f\alpha^f_{1}\partial_{\mu}\nu_{f,\nu}\right)\nonumber\\
&&-\sum_f\frac{\nu_f^{\mu}}{T}\left[T\partial_{\mu}\left(\frac{\mu_f}{T}\right)+q_fF_{\mu\nu}u^{\nu}-\sum_{f'}\beta^{ff'}_1\dot\nu_{f',\mu}
+\alpha^f_{0}\partial_{\mu}\Pi+\alpha^f_{1}\partial_{\nu}\pi^{\nu}_{\mu}\right],
\end{eqnarray}
where $\dot f\equiv u^{\mu}\partial_{\mu}f$.
Constitutive equations which ensure the second-law of thermodynamics are obtained as follows
\begin{eqnarray}
-\Pi&=&\zeta\left(\triangle^{\mu\nu}\partial_{\mu}u_{\nu}+\beta_0\dot\Pi+\sum_f\alpha^f_{0}\partial_{\mu}\nu_f^{\mu}\right),\\
\pi^{\mu\nu}&=&2\eta\left\langle\left\langle\partial^{\mu}u^{\nu}-\beta_2\dot\pi^{\mu\nu}-\sum_f\alpha^f_{1}\partial^{\mu}\nu_{f}^{\nu}\right\rangle\right\rangle,\\
\nu_f^{\mu}&=&\sigma_f \triangle^{\mu}_{\rho}\sum_{f'}\kappa^{ff'}\left[T\partial^{\rho}\left(\frac{\mu_{f'}}{T}\right)+q_{f'}F^{\rho}_{\sigma}u^{\sigma}-\sum_{f''}\beta^{f'f''}_1\dot\nu_{f''}^{\rho}+\alpha^{f'}_{0}\partial^{\rho}\Pi+\alpha^{f'}_{1}\partial_{\sigma}\pi^{\sigma\rho}\right],
\end{eqnarray}
with bulk and shear viscosities $\zeta, \ \eta \ (\geq0)$, flavor conductivity $\sigma_f \ (\geq 0)$, and a positive semi-definite flavor mixing matrix $\kappa^{ff'}$.
$\langle\langle B^{\mu\nu}\rangle\rangle$ stands for
\begin{eqnarray}
\langle\langle B^{\mu\nu}\rangle\rangle\
\equiv \triangle^{\mu\rho}\triangle^{\nu\sigma}
\left[\frac{B_{\rho\sigma}+B_{\sigma\rho}}{2}-\frac{\triangle_{\rho\sigma}\triangle^{\alpha\beta}B_{\alpha\beta}}{3}\right].
\end{eqnarray}

We perform linear analysis in terms of $\delta e,\delta n_f$, and $\delta \vec u$ defined as
\begin{eqnarray}
e(x)&=&e+\delta e(x),\\
n_f(x)&=&n_f+\delta n_f(x),\\
u^{\mu}(x)&=&(1,\ \delta \vec u(x)),
\end{eqnarray}
and external vector field $\delta A^{\mu}$.
For simplicity we neglect the couplings between different dissipative terms: $\alpha^f_{0}=\alpha^f_{1}=0$, $\beta^{ff'}_1=\beta^{f}\delta_{ff'}$, and $\kappa^{ff'}=\kappa^f\delta_{ff'}$ (assumption (i)).
The dissipative flavor current $\nu_f^{\mu}(x)$ is then given by
\begin{eqnarray}
\nu_f^0(x)&=&0, \\
\vec\nu_f(x)&=&-\sigma_f\left[
T\vec\nabla\left(\frac{\mu_f}{T}\right)
-q_f\vec E+\beta^f_{1}\partial_t\vec\nu_f
\right],
\end{eqnarray}
in the linear order in $\delta e$, $\delta n_f$, $\delta \vec u$, and $A^{\mu}$.
We define susceptibility matrices $X$ and $X_{\rm f}$ as
\begin{eqnarray}
X\equiv\begin{pmatrix}
\frac{\partial e}{\partial T} & \frac{\partial e}{\partial \mu_{\rm u}} & \frac{\partial e}{\partial \mu_{\rm d}} & \frac{\partial e}{\partial \mu_{\rm s}} \\
\frac{\partial n_{\rm u}}{\partial T} & \frac{\partial n_{\rm u}}{\partial \mu_{\rm u}} & \frac{\partial n_{\rm u}}{\partial \mu_{\rm d}} & \frac{\partial n_{\rm u}}{\partial \mu_{\rm s}} \\
\frac{\partial n_{\rm d}}{\partial T} & \frac{\partial n_{\rm d}}{\partial \mu_{\rm u}} & \frac{\partial n_{\rm d}}{\partial \mu_{\rm d}} & \frac{\partial n_{\rm d}}{\partial \mu_{\rm s}} \\
\frac{\partial n_{\rm s}}{\partial T} & \frac{\partial n_{\rm s}}{\partial \mu_{\rm u}} & \frac{\partial n_{\rm s}}{\partial \mu_{\rm d}} & \frac{\partial n_{\rm s}}{\partial \mu_{\rm s}} 
\end{pmatrix}, \ \ \ 
X_{\rm f}\equiv\begin{pmatrix}
\frac{\partial n_{\rm u}}{\partial \mu_{\rm u}} & \frac{\partial n_{\rm u}}{\partial \mu_{\rm d}} & \frac{\partial n_{\rm u}}{\partial \mu_{\rm s}} \\
\frac{\partial n_{\rm d}}{\partial \mu_{\rm u}} & \frac{\partial n_{\rm d}}{\partial \mu_{\rm d}} & \frac{\partial n_{\rm d}}{\partial \mu_{\rm s}} \\
\frac{\partial n_{\rm s}}{\partial \mu_{\rm u}} & \frac{\partial n_{\rm s}}{\partial \mu_{\rm d}} & \frac{\partial n_{\rm s}}{\partial \mu_{\rm s}} 
\end{pmatrix}.
\end{eqnarray}
Then the dissipative flavor current $\nu_f^{\mu}$ is obtained as
\begin{eqnarray}
\nu_f^0&=&0 , \ \ \ 
\vec \nu_f \ = \ q_f\sigma_f\vec E-\beta_{f,1}\sigma_f\partial_t\vec \nu_f
-\sigma_f\left(-\frac{\mu_f}{T},\left\{{\vec {\rm e}_f}\right\}\right)X^{-1}
\begin{pmatrix}
\vec\nabla\delta e \\
\vec\nabla\delta n_{\rm u} \\
\vec\nabla\delta n_{\rm d} \\
\vec\nabla\delta n_{\rm s} 
\end{pmatrix},\\
\left\{\vec{\rm e}_{\rm u}\right\}&\equiv &\{1,0,0\} , \ \ \
\left\{\vec{\rm e}_{\rm d}\right\} \ \equiv \ \{0,1,0\} , \ \ \
\left\{\vec{\rm e}_{\rm s}\right\} \ \equiv \ \{0,0,1\} .
\end{eqnarray}

We make another simplification that $\sigma_f$ and $\beta^f_{1}$ are flavor independent (assumption (ii)):
\begin{eqnarray}
\sigma_f\equiv\bar\sigma, \ \
\beta^f_{1}\equiv\bar\beta_1,
\end{eqnarray}
and restrict ourselves to the situation with vanishing flavor chemical potential $\mu_f=0$, which decouples flavor dissipation and sound mode propagation.
We then arrive at
\begin{eqnarray}
\vec\nu_f+\bar\beta_1\bar\sigma\frac{\partial}{\partial t}\vec \nu_f=q_f\bar\sigma\vec E-\bar\sigma\left\{\vec{\rm e}_f\right\} X^{-1}_{\rm f}
\begin{pmatrix}
\vec\nabla\delta n_{\rm u} \\ \vec\nabla\delta n_{\rm d} \\ \vec\nabla\delta n_{\rm s}
\end{pmatrix}.
\end{eqnarray}

So far we have simplified transport coefficients in order to reduce unknown parameters and to avoid mode couplings.
In order to obtain simple constitutive equation for diffusive electric current in multi-flavor case, it is necessary to simplify thermodynamic quantities, namely susceptibility matrix to be $X_{\rm f}\approx\left(\partial n_{\rm u}/\partial\mu_{\rm u}\right){\bf 1}\equiv \chi_{\rm uu}{\bf 1}$ (assumption (iii)).
This yields
\begin{eqnarray}
\vec\nu_f+\bar\beta_1\bar\sigma\frac{\partial}{\partial t}\vec \nu_f=q_f\bar\sigma\vec E-\frac{\bar\sigma}{\chi_{\rm uu}}\vec\nabla\delta n_f.
\end{eqnarray}
Note that by this simplification we also obtain electric charge susceptibility $\chi$ in terms of $\rm u$-quark number susceptibility $\chi=(2/3)\chi_{\rm uu}$.
Then constitutive equation for diffusive electric current is given by
\begin{eqnarray}
\label{mSPF:eq:constitutive_linear}
\delta n\equiv\sum_fq_f\delta n_f, \ \ \ \vec\nu\equiv\sum_fq_f\vec\nu_f, \ \ \ 
\vec\nu+\bar\beta_1\bar\sigma\frac{\partial}{\partial t}\vec\nu=\frac{2}{3}\bar\sigma\vec E-\frac{2\bar\sigma}{3\chi}\vec\nabla\delta n.
\end{eqnarray}
Comparing with Eq.~(\ref{DL:eq:constitutive_linear}), transport coefficients for the electric current are related to those for flavor currents by
\begin{eqnarray}
\tau_{\rm J}=\bar\beta_1\bar\sigma, \ \ \ 
\sigma=\frac{2}{3}\bar\sigma, \ \ \ 
D=\frac{2\bar\sigma}{3\chi},
\end{eqnarray}
satisfying $\sigma=\chi D$.

We are also interested in the constitutive equation for diffusive quark number current.
Then we need to modify Eq.~(\ref{mSPF:eq:constitutive_linear}) by substituting $q_f\rightarrow 1, \delta A^{\mu}\rightarrow \delta A^{\mu}_{\rm q}$, and $\vec E\rightarrow \vec E_{\rm q}$, where $\delta A^{\mu}_{\rm q}$ denotes external vector field that couples with quark number current.
Note also that total quark number susceptibility is $\chi_{\rm q}=3\chi_{\rm uu}$ because of the simplification for $X_{\rm f}$.
Then we get
\begin{eqnarray}
&&\delta n_{\rm q}\equiv\sum_f\delta n_f, \ \ \ \vec\nu_{\rm q}\equiv\sum_f\vec\nu_f, \ \ \ 
\vec\nu_{\rm q}+\bar\beta_1\bar\sigma\frac{\partial}{\partial t}\vec\nu_{\rm q}=3\bar\sigma\vec E_{\rm q}-\frac{3\bar\sigma}{\chi_{\rm q}}\vec\nabla\delta n_{\rm q},\\
&&\tau_{\rm q}=\bar\beta_1\bar\sigma, \ \ \ 
\sigma_{\rm q}=3\bar\sigma, \ \ \
D_{\rm q}=\frac{3\bar\sigma}{\chi_{\rm q}},
\end{eqnarray}
satisfying $\sigma_{\rm q}=\chi_{\rm q}D_{\rm q}$.
Therefore we can verify $D=D_{\rm q}$, which we use in the text.

\newpage
\bibliographystyle{apsrev}

\begin{thebibliography}{99}
\bibitem{footnote1}
Since we are interested in the quark-gluon plasma at vanishing chemical potential, the Eckart frame, in which we define the flow vector by $u^{\mu}$ by $J^{\mu}=\rho u^{\mu}$, is inconvenient.
This is because the electric current is $J^{\mu}=(0,0,0,0)$ in equilibrium and hence $u^{\mu}$ becomes ill-defined.
\bibitem{footnote2}
We can also parameterize $\tau_{\rm J}\propto 1/T$ and $\sigma\propto T$ assuming the scaling to temperature.
This parameterization differs from the one adopted in our study only below the transition region $T<T^*_{\rm c}$ if we fix the higher temperature ($T\sim 2T^*_{\rm c}$) behavior.
\end{thebibliography}

\end{document}